\documentclass[manuscript,screen]{acmart}
\usepackage[ruled,linesnumbered]{algorithm2e}

\SetCommentSty{mycommfont}
\usepackage{graphicx}
\usepackage{textcomp}
\usepackage{multirow}
\usepackage[normalem]{ulem}
\usepackage{xcolor}
\usepackage{booktabs}
\usepackage{tabularx}
\usepackage{colortbl}
\usepackage{array}
\usepackage{enumitem}

\AtBeginDocument{%
  \providecommand\BibTeX{{%
    \normalfont B\kern-0.5em{\scshape i\kern-0.25em b}\kern-0.8em\TeX}}}

\newcommand{\sw}[1]{\textcolor{purple}{{\it [Shaowei says: #1]}}}

\begin{document}

%%
%% The "title" command has an optional parameter,
%% allowing the author to define a "short title" to be used in page headers.
\title{SimClone: Detecting Tabular Data Clones using Value Similarity}

%%
%% The "author" command and its associated commands are used to define
%% the authors and their affiliations.
%% Of note is the shared affiliation of the first two authors, and the
%% "authornote" and "authornotemark" commands
%% used to denote shared contribution to the research.
\author{Xu Yang}
% \authornote{Both authors contributed equally to this research.}
\email{yangx4@myumanitoba.ca}
% \orcid{1234-5678-9012}
\affiliation{%
  \institution{University of Manitoba}
 % \streetaddress{66 Chancellors Circle}
  %\city{Winnipeg}
  %\state{Manitoba}
  \country{Canada}
  % \postcode{43017-6221}
}

\author{Gopi Krishnan Rajbahadur}
\affiliation{%
  \institution{Centre for Software Excellence, Huawei Canada}
  % \streetaddress{1 Th{\o}rv{\"a}ld Circle}
  % \city{Hekla}
  \country{Canada}}
\email{gopi.krishnan.rajbahadur1@huawei.com}

\author{Dayi Lin}
\affiliation{%
  \institution{Centre for Software Excellence, Huawei Canada}
  % \city{Rocquencourt}
  \country{Canada}
}
\email{dayi.lin@huawei.com}

\author{Shaowei Wang}
\email{shaowei.wang@umanitoba.ca}
\affiliation{%
\institution{University of Manitoba}
 % \streetaddress{66 Chancellors Circle}
  %\city{Winnipeg}
  %\state{Manitoba}
  \country{Canada}
}

\author{Zhen Ming (Jack) Jiang}
\email{zmjiang@cse.yorku.ca}
\affiliation{%
\institution{York University}
  % \streetaddress{66 Chancellors Circle}
  % \city{Winnipeg}
  % \state{Manitoba}
  \country{Canada}
}

%%
%% By default, the full list of authors will be used in the page
%% headers. Often, this list is too long, and will overlap
%% other information printed in the page headers. This command allows
%% the author to define a more concise list
%% of authors' names for this purpose.
\renewcommand{\shortauthors}{Xu et al.}

%%
%% The abstract is a short summary of the work to be presented in the
%% article.
\begin{abstract}
Data clones are defined as multiple copies of the same data among datasets. Presence of data clones between datasets can cause issues such as difficulties in managing data assets and data license violations when using datasets with clones to build AI software. However, detecting data clones is not trivial. Majority of the prior studies in this area rely on structural information to detect data clones (e.g., font size, column header). However, tabular datasets used to build AI software are typically stored without any structural information. In this paper, we propose a novel method called SimClone for data clone detection in tabular datasets without relying on structural information. SimClone method utilizes value similarities for data clone detection. We also propose a visualization approach as a part of our SimClone method to help locate the exact position of the cloned data between a dataset pair. Our results show that our SimClone outperforms the current state-of-the-art method by at least 20\% in terms of both F1-score and AUC. In addition, SimClone's visualization component helps identify the exact location of the data clone in a dataset with a Precision@10 value of 0.80 in the top 20 true positive predictions.
\end{abstract}

%%
%% The code below is generated by the tool at http://dl.acm.org/ccs.cfm.
%% Please copy and paste the code instead of the example below.
%%


%%
%% Keywords. The author(s) should pick words that accurately describe
%% the work being presented. Separate the keywords with commas.

%%
%% This command processes the author and affiliation and title
%% information and builds the first part of the formatted document.
\maketitle

\section{Introduction}
Datasets form a key component in the development of Artificial Intelligence (AI) software, whose adoption and commercialization have increased significantly in the past decade~\cite{gartner2019}. These datasets are commonly created by combining multiple datasets, scraping several sources, or merging different tables in a data warehouse~\cite{Gopi2021public,benjamin2019towards}. For example, CIFAR-10, a commonly used dataset to build AI software, was created by extracting a subset of images from the \emph{80 Million Tiny Images} dataset, which was in turn created from several different data sources including Google Images, Flickr, and Ask~\cite{krizhevsky2009learning}. Such practices create data clones among datasets.

% why detecting data clone is important
%\sw{R2.1}
We define data clones as multiple copies of identical data across datasets. Data clones presented in the dataset could raise concerns and risks for AI developers and companies from various aspects. 
First, many datasets have specific usage restrictions or requirements for attribution. Misusing or abusing such data could lead to copyright infringement and license issues~\cite{birhane2021large,xia2023empirical}. Second, data clones could lead to data leakage and introduce bias for AI model training and evaluation~\cite{jin2021cafe,shabtai2012data,alneyadi2016survey}. For instance, if certain types of data are overrepresented due to cloning, the resulting models may exhibit skewed predictions or decisions, favoring the duplicated data at the expense of underrepresented or minority groups. This can exacerbate existing biases in the data, leading to unfair outcomes in AI applications~\cite{drukker2023toward,codevilla2019exploring,allamanis2019adverse}.
Therefore, in the realm of software engineering, the detection and management of data clones are of paramount importance. It enables AI software developers to maintain data integrity, ensuring traceability and compliance with data provenance and licensing requirements. 

%Being able to detect the presence of data clones among datasets is important for AI software developers as it allows them to avoid potential dataset license violations, helps them achieve better traceability in the development of AI software, and enables them to verify data provenance. For example, the current license of \emph{80 Million Tiny Images} dataset states that the dataset should not be used for AI software development since the data might contain obscene images~\cite{birhane2021large,xia2023empirical}. Therefore, being able to detect that CIFAR-10 contains data clones from the \emph{80 Million Tiny Images} dataset could help AI software developers avoid or mitigate the risk of potential license violations. In addition, governments around the world are mandating that the origin of code, data, and other artifacts are recorded using standard documents, i.e., Software Bill Of Materials (SBOM)~\cite{white2021}. Being able to detect data clones among datasets and data sources enables AI software developers to recover, document, and verify the provenance of data points contained in a dataset.

Although prior studies~\cite{popescu2004exposing,wang2006large,tamilselvi2011handling,Hermans2013,Dou2016,zhang2020learning,alotaibi2017etdc} have proposed several methods that tackle related problems, there are several shortcomings when applying such methods to detect data clones in the above-mentioned context. First, many proposed methods in the literature can only detect data clones in homogeneous datasets (e.g., image datasets~\cite{popescu2004exposing, wang2006large}), not structured heterogeneous datasets (e.g., tabular datasets)~\cite{borisov2022deep}, despite tabular data being the most common data type in real-world AI applications~\cite{shwartz2022tabular}. Second, for methods that work on tabular datasets, they either only consider record-level clones and ignore column-level clones (commonly referred to as data duplication detection)~\cite{alotaibi2017etdc,tamilselvi2011handling,10.1145/3310205}, or heavily rely on structural or formatting information such as row or column headers, formulas, and background colors~\cite{Hermans2013,Dou2016,zhang2020learning}, which are not available in tabular datasets that are used in AI software development. Tabular datasets used in AI software development are usually large and value-oriented. Therefore, existing approaches are not applicable to tabular datasets for real-world AI software development. %The data duplicate detection only considers record level duplicates and for a data clone detection to be effective both record level and column level clones need to be detected. For tabular data clone detection approaches they typically rely on the formatting information provided in the spreadsheets or rules. For instance, a recent rule-based method TableCheck~\cite{Dou2016} works by detecting if two values have same row or column headers which will not detect clones following row or column shifts. Feature-based methods like LTC (Learning to Detect Table Clones)~\cite{zhang2020learning} works by analyzing if a two datasets contain similar structural information like headers, formulas and background colors. These methods may not be robust to variations in the data and can be constrained by the availability of features in tabular datasets. However, the tabular datasets used to build AI Software typically are stored in serialized formats like pickle or simple formats like csv which lack such formatting or structural information.  

In this paper, we propose a novel method titled ``Value-based \textbf{Sim}ilarity for Data \textbf{Clone} Detection'' (SimClone) to detect data clones among tabular datasets. SimClone overcomes the limitations of prior work by requiring no format-related metadata. SimClone is based on the premise that values within tabular datasets represent the most discriminative attribute of clones and that the presence of similar values between two datasets is indicative of the existence of data clones. To detect data clones, SimClone first computes 14 value similarity features using 6 value-based similarity metrics (i.e., Jaccard~\cite{jaccard}, Textrank~\cite{mihalcea2004textrank}, Simhash~\cite{simhash}, Levenshtein~\cite{Levenshtein}, Mean and Standard Deviation) between each dataset pair, and then uses a supervised machine learning classifier to detect if there are data clones between the two paired datasets. Additionally, we propose a visualization method that enables SimClone users to locate the exact position of cloned data between a dataset pair by combining the feature importance scores that we obtain by interpreting the constructed classifier using SHAP~\cite{shap} with the similarity matrices obtained in the previous steps. The classifier is trained on 11,935 pairs of datasets synthesized from injecting clones into 154 datasets in the UCI machine learning repository. We evaluate SimClone against the state-of-the-art tabular data clone detection method LTC~\cite{zhang2020learning} on both the synthesized test set and real-world datasets EUSES~\cite{Fisher2005}. 

SimClone outperforms the state-of-the-art (SOTA) method LTC in all evaluation metrics. For instance, SimClone achieves an F1 of 0.83, improving LTC by at least 32.3\% on the synthetic test set. SimClone improves LTC at least 100\% on the real-world EUSES dataset in terms of Precision@K even when K=200. In addition, our visualization method can improve the baseline by at least 45\% in locating the positions of cloned data in the detected clone pairs. %In addition, we discuss several ways of increasing the performance of our SimClone method and discuss the implications of having a tool such as our for the AI software developers. 
In addition, we make the code for SimClone and our experiments (including the synthetic datasets that we created) openly available in a replication package~\cite{simclonerepo} to foster open science and provide a benchmark for tabular data clone detection that can be used by future research.

%Overall, our proposed SimClone approach effectively detects data clones in tabular datasets, while also providing interpretability and transparency in the process.

%The results of our experiment also demonstrated that the SimClone outperformed the current state-of-the-art table clone detection algorithm, LTC, by more than 20\% on f1 and recall metrics when evaluated on the UCI Machine Learning dataset repository.

In summary, the contribution of our paper includes:
\begin{itemize}
   \item To the best of our knowledge, we are the first to introduce value similarity metrics into the detection of data clones between tabular datasets. Our study provides a framework for use in real-world production environments and offers insight for future research on data clones.
   \item We conducted experiments on the UCI Machine Learning dataset repository and also the real-world dataset EUSES and Enron. Our experimental results suggest promising performance in detecting data clones among tabular datasets. We open-sourced our data and code implementation\cite{simclonerepo}.
   \item We proposed a visualization method that leverages interpretable AI technique SHAP~\cite{shap} to facilitate the visualization process of SimClone, our experiment result suggests the methods can help users better locate the data clone. The idea of combining interpretable AI with visualization could inspire future research.
   \item We have released a clone-injected subset of the UCI Machine Learning dataset repository as a benchmark for the detection of data clones in tabular datasets for future studies.
\end{itemize}

%\sw{remove this paragraph if need more space}
\smallskip\noindent\textbf{Paper organization.} Section~\ref{sec:background} defines different types of data clones in tabular datasets. Section~\ref{sec:related} presents the related work and Section~\ref{methodology} provides the methodology of SimClone. In Section~\ref{sec:exp}, we describe the experiment design that is used to evaluate SimClone. In Section~\ref{sec:results}, we present the results of our RQs. In Section~\ref{sec:discussion}, we discuss the implications of our study and point out the future direction. In Section~\ref{sec:threats}, we provide the threats to the validity of our study. Finally, we conclude our paper in Section~\ref{sec:conclusion}.

\section{Definition of Data Clone}\label{sec:background}
\label{sec:background}
%\subsection{Data Clone in Tabular Datasets}
\begin{table*}[]
\caption{General definition of data clone and example cases for tabular datasets}
% \sw{I guess we can have a figure to demo each type of data clone to help reader understand the concept.}
\begin{tabularx}{\textwidth}{|ll|X|X|}
\cline{1-4}
\multicolumn{2}{|l|}{} &
  \textbf{General Definitions of Data Clone} &
  \textbf{Example Cases of Clones among Tabular Datasets} \\ \hline
\multicolumn{2}{|l|}{\multirow{2}{*}{\textbf{Terms}}} &
  Data records &
  Rows/columns \\ \cline{3-4} 
\multicolumn{2}{|l|}{} &
  Data items &
  Cells \\ \hline
\multicolumn{1}{|c|}{\multirow{4}{*}{\textbf{Clone Types}}} &
  1 &
  Exact match of data records or data items (only difference in metadata such as modified date, file name, etc.). &
  A consecutive group of identical cell values. \\ \cline{2-4} 
\multicolumn{1}{|c|}{} &
  2 &
  Exact content of data records or data items with differences in format. &
  Numeric values with different precisions; textual values with different encodings.\\ \cline{2-4} 
\multicolumn{1}{|c|}{} &
  3 &
  Simple transformation of data records or data items. &
  Adding/dropping a certain amount of columns or rows in tabular data, switching the order of columns or rows. \\ \cline{2-4} 
\multicolumn{1}{|c|}{} &
  4 &
  Complex transformation of data records or data items which are derivative of original data records or data items. &
  Values that have been transformed using differential privacy techniques (e.g., adding small noise to the data). \\ \cline{1-4} 
\end{tabularx}
\label{tab:def}
\end{table*}

%\begin{figure}[ht]	
%    \centering
%    \includegraphics[width=\columnwidth]{Figures/tabular_clone_example.png}
%    \caption{data clone example in tabular datasets}
%    \label{tab:data_clone_example}
%\end{figure}
% Previous studies have conducted research on various forms of data clones and have provided different definitions. For example, in the context of spreadsheet data clones, they are generally referred to as groups of cell blocks that share the same text, value, or formula. In the context of programming code data clones, they are considered to be two snippets of code that are semantically or logically identical. However, no previous studies have focused on data clones in tabular datasets, and therefore, a definition of data cloning in tabular datasets has not yet been established. Therefore, we have attempted to provide a definition of data clones in tabular datasets based on the characteristics of tabular datasets and in reference to other types of data.
In previous studies, the concept of data clone is defined mostly in the context of spreadsheets, as cell blocks that are created by directly copy-pasting values or computational semantics (i.e., formula)~\cite{Dou2016, Hermans2013}. The main motivation behind defining data clones in such a context is that users may introduce errors in the data when copy-pasting cell blocks, especially when the cells contain formulas. However, such definitions do not cover issues specific to the usage of data in AI software development (e.g., traceability, provenance, and data license conformance). In particular, most tabular datasets used for machine learning are plain values without formulas. In addition, a data license may also prohibit certain usages of any derivatives from the original data (i.e., data transformation), which is beyond the scope of the spreadsheet data clone definitions but overseeing such clones may result in data license violation. Finally, such definitions do not consider other unstructured data types such as images. %In the context of programming code data clones, they are considered to be two or more snippets of code that are semantically or logically identical~\cite{Shobha2021CodeCD}. However, these definitions are inadequate for understanding data clones in tabular datasets \sw{again what is the difference between tabular and table}\xy{good questions I dont know also...@Gopi}, as they have not been previously studied in this context. For example, a table might be used to display a summary of data in a report or presentation, while a tabular dataset would be used for data analysis, machine learning, and other computational tasks, which result in a different volume of data and a different structural.

Hence, we propose a set of general definitions for different types of data clones, that are not limited to the context of copy-pasting in spreadsheets, and can be extended to more data types. Table~\ref{tab:def} presents our general definitions and examples of cases of clones in tabular datasets under each type. Inspired by the definitions of different types of code clones, we also use different types to describe different levels of data clones. The definition takes into account the unique characteristics of tabular datasets, specifically Translation Invariance and Value Orientation. Translation Invariance refers to the property of a tabular dataset that allows for it to be considered the same, regardless of shifts or changes in column or row positions. Value Orientation pertains to the unique characteristics of the data being the most distinct attributes of the dataset. 

Our definitions are to be distinguished from another relevant but different concept - data duplication. Under our definition, data clones in tabular datasets can occur in two dimensions: both rows and columns, whereas data duplication only concerns repetition at the row (i.e., record) level. Hence data duplication can be considered a special case of data cloning under our definitions.

In this study, we mainly focus on detecting Type-1 and Type-3 data clones among tabular datasets. Our method also works for Type-2 data clones on numerical values.
% In tabular datasets, data clones may occur in two dimensions. First, there may be data clones of rows, and since a row of a tabular dataset typically represents a data point, data clones of rows also imply the cloning of data points. Second, there may also be data clones of columns in a tabular dataset, where a column in a tabular dataset typically represents a feature. In this study, we aim to detect the Type-1 to Type-3 data clone among tabular datasets.

\section{Related Work}\label{sec:related}
\label{sec:related}
In this section, we discuss related work to our study. %There are three research areas that are related to detecting or deleting ``clone'' in data, and inspired our study. We review these approaches in this section.
\subsection{Data quality issues}
%\sw{R2.2}
Modern artificial intelligence (AI) applications require a large amount of training and test data, which creates critical challenges not only concerning the availability of such data, but also regarding its quality. For example, data quality related to incomplete, inconsistent, dated, duplicated, or inappropriate training data can lead to unreliable models that produce ultimately poor decisions~\cite{batini2009methodologies,gudivada2017data,jain2020overview,zhou2017machine}. There are remarkable efforts that have been invested in improving the data quality from various aspects~\cite{fenza2021data,cong2007improving,pleiss2020identifying,frenay2013classification}. For instance, to improve the data consistency and accuracy, Gao et al.~\cite{cong2007improving} proposed an approach to identify the data consistency and accuracy by using conditional functional dependencies and also developed algorithms to fix the data quality issues. Pleiss et al.~\cite{pleiss2020identifying} proposed a method to identify mislabeled data and mitigate their impact when training neural networks by using the Area Under the Margin (AUM) statistic, which
exploits differences in the training dynamics of clean and mislabeled samples. 

Data duplication is the most related data quality issue to our study. If certain types of data are overrepresented due to duplication, the resultant models may exhibit skewed predictions or decisions, favoring the duplicated data at the expense of underrepresented or minority groups~\cite{allamanis2019adverse}. 
Data deduplication is widely researched in the data management community. Numerous approaches for data deduplication have been proposed. As summarized by Ilyas et al.~\cite{10.1145/3310205}, data deduplication methods can be classified into two categories: unsupervised and supervised. Unsupervised methods, such as those based on a pre-specified threshold~\cite{monge1996field,chaudhuri2005robust}, or domain-specific rules~\cite{hernandez1998real,doan2003profile,weis2008industry}, do not rely on labeled clone pairs. On the contrary, supervised methods such as Naive Bayes~\cite{winkler1999state}, decision trees~\cite{chaudhuri2007example}, or support vector machines~\cite{bilenko2003adaptive} require labeled clone pairs to train models. In recent years, there has been a growing interest in the role of humans in the data deduplication process to improve the accuracy of automatic data deduplication. Wang et al. proposed CrowdER~\cite{wang2012crowder}, which only sends high-probability clone pairs identified by machine learning models to be verified by crowd-sourced platforms. Gokhale et al. proposed Corleone~\cite{gokhale2014corleone}, which uses a combination of blocking rules and active learning to improve accuracy while minimizing crowdsourcing costs.

As mentioned in Section~\ref{sec:background}, data duplications are a special case of data clones, where only record-level clones (i.e., rows in tabular datasets) are considered. In other words, data clone is generalized from data duplication and is also a type of data quality issue. 
However, due to the difference between data clone and data duplication, existing data deduplication methods are not applicable in our context. Therefore, in this study, we develop SimClone which is designed to identify data clones in tabular datasets.

\subsection{Data Clone Detection in Spreadsheets}
Spreadsheets are one format of tabular datasets. Prior work on clones in tabular datasets mostly focuses on copy-pasting across spreadsheets, because such action can introduce errors especially when formulas are presented. %However, the key difference between the two is the format and organization of the data. Spreadsheet tables are organized in a grid-like format with rows and columns, and the data is typically arranged in a tabular format, with a fixed structure and additional features like formulas, and charts. Tabular datasets, on the other hand, can be stored in a variety of formats such as CSV, JSON, or SQL, have a flexible structure and do not include additional features. 
Hermans et al.~\cite{Hermans2013} showed that data clones are common among spreadsheets and pose a threat to spreadsheet quality, and proposed an algorithm to identify and resolve clones. Dou et al.~\cite{Dou2016} proposed TableCheck, a tool that detects spreadsheet data clones based on the observation that two tables with the same header information are likely to be cloned. Zhang et al.~\cite{zhang2020learning} proposed LTC (Learning to detect Table Clones), which uses information such as row header name, font type, and font color, among other format features to detect spreadsheet data clones. 

As discussed in Section~\ref{sec:background}, tabular datasets that are used for AI software development do not contain most format information. In addition, many existing methods (e.g., TableCheck, LTC) cannot identify cloned data blocks that have gone through simple transformations such as reshuffled row and column orders (i.e., Type 3 data clone). Therefore, existing spreadsheet data clone detection methods cannot be directly applied in the context of AI software development.

\subsection{Detecting Clones and Duplicates in other Software Artifacts}
Clone as a concept has been extensively studied in the software engineering community for various artifact types, such as code clones and bug report duplicates. For instance, many methods have been proposed for code clone detection, including lexical analysis, Abstract Syntax Tree (AST) comparison, metric-based, program slicing, and machine learning~\cite{ain2019systematic,shobha2021code,komondoor2001slice}. In addition, prior work has also explored detecting clones in images~\cite{popescu2004exposing, wang2006large}. 

%\sw{R2.9}
Recently, with the rise of pre-trained models and their state-of-the-art performance in various software engineering (SE) tasks, they have been increasingly applied to detect clones and duplicates in software artifacts~\cite{feng2020codebert,guo2020graphcodebert,tao2022c4,isotani2021duplicate}. For example, Feng et al. introduced CodeBERT, a pre-trained model based on the BERT architecture specifically trained on source code for code-related tasks~\cite{feng2020codebert}. They demonstrated its effectiveness in capturing source code semantics and measuring their similarity in code detection tasks. Similarly, Guo et al. introduced Graphcodebert~\cite{guo2020graphcodebert}, which incorporates structural information (e.g., data flow) in training and achieves state-of-the-art performance in code clone detection. Isotani et al. utilized BERT to detect duplicate bug reports~\cite{isotani2021duplicate}. 

However, due to the nature of tabular data, which consists of a mixture of numeric and string values, directly applying such pre-trained models is challenging. For instance, it is difficult to directly apply a pre-trained model like BERT to measure the similarity of two sets of numeric values. Therefore, we propose our value-based similarity approach to individually measure the similarity between string and numeric values. This approach enables more effective comparison and identification of similarities within tabular data.

\section{Methodology}\label{methodology}
\begin{figure*}[ht]	
    \includegraphics[width=0.9\textwidth]{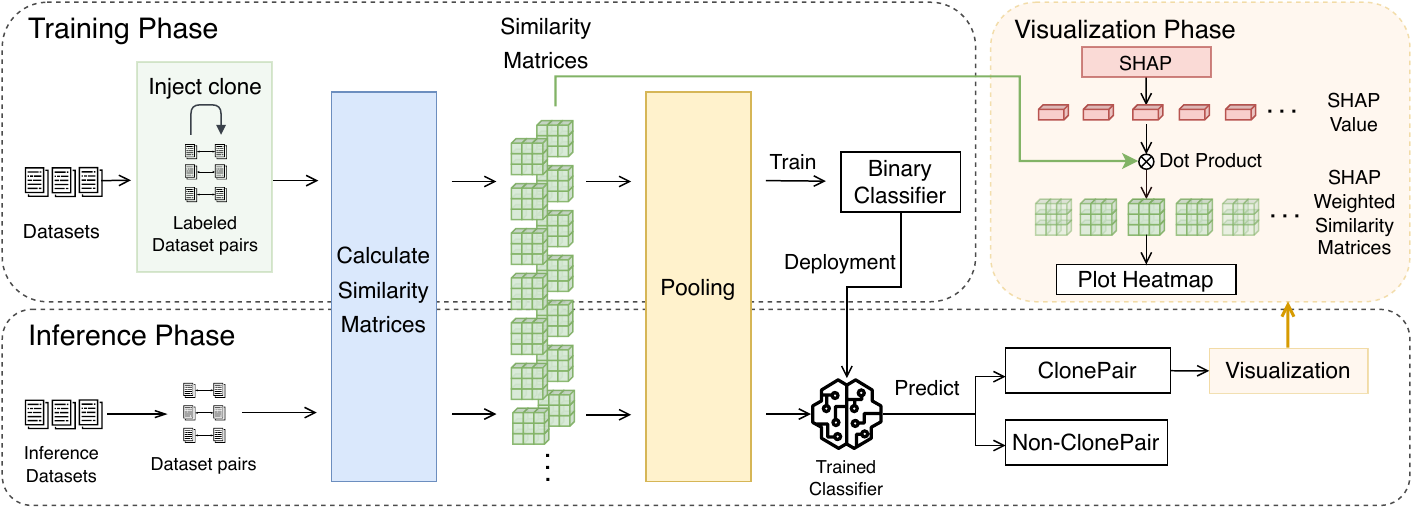}
    \caption{Overview of SimClone.}
    \label{fig:pipeline}
\end{figure*}

In this section, we introduce our proposed value-based \textbf{Sim}ilarity for data \textbf{Clone} Detection (SimClone) method in detail.

We define the task of data clone detection as follows: 
Given a set of $n$ datasets $Set_{dataset}= \{D_1..D_n\}$, we aim to find all dataset pairs $\langle D_i, D_j\rangle \in Set_{dataset}$ that contain data clones (referred to as $ClonePair$) and indicate the location of data clone in each $ClonePair$. For simplicity, we use dataset and tabular dataset exchangeably in the following sections. 
%\sw{R2.4}
However, identifying data clones is a challenging task. There are several challenges for data clone detection. Firstly, we focus on Type-1, Type-2, and Type-3 clones. There could be various forms and sizes of clones in each clone type. For instance, Type-3 data clones focus on simple transformation of data records or data items, such as adding/dropping a certain amount of columns or rows in tabular data, or switching the order of data. There are infinite ways to transform the tabular data from one form into another form with those simple transformations. In addition, the clone could be either text, numerical values, or their combination, we need to design different features to capture their characteristics and measure their similarity. Therefore, it is challenging to manually craft hard rules (e.g., deciding the threshold for similarity, assigning weight for different features, etc.) to identify data clones with various transformations and forms. Even though manually crafting hard rules is possible, it is a labor-intensive process.
Second, even if a pair of tabular data is identified as a \textit{ClonePair}, with enormous data points in the pair, it is still a challenging job to identify the exact location of the cloned data.
Therefore, to address the above challenges, we propose SimClone to identify \textit{ClonePairs} by learning a classifier automatically from data clones and visualize the exact location of cloned data within a \textit{ClonePair} to help practitioners locate the clone data. 
Training a classifier requires a large amount of data, however, no existing data clone dataset could leveraged for this purpose. Therefore, We first create a synthetic dataset where we artificially inject Type-1 and Type-3 data clones. We then pair every dataset with every other dataset in the synthetic dataset and label which pairs are \textit{ClonePairs} among them. We derive value similarity-based features between each dataset pair and train a binary classifier on this synthetic dataset to identify \textit{ClonePairs}. We leverage this trained classifier to identify \textit{ClonePairs} among a given set of dataset pairs. In addition, we develop a visualization approach to indicate the exact location of the cloned data within a \textit{ClonePair}. Thus, for SimClone, the input is a pair of two datasets and the output is a binary classification result (e.g. whether the given dataset pair is a \textit{ClonePair}), plus a visualization result indicating the location of the data clone if the dataset pair is a \textit{ClonePair}. Figure~\ref{fig:pipeline} presents an overview of our SimClone method, which is composed of training and inference phases. We elaborate on the details of each step of SimClone in the following subsections.

\subsection{Synthetic dataset creation}\label{sec:exp_data}

For our SimClone method, we need a dataset comprised of tabular dataset pairs with labels indicating if they are a \textit{ClonePair}. However, as far as we know, such a dataset with labeled \textit{ClonePairs} is not readily available. Hence, we create a collection of synthetic tabular datasets that contain data clones for our experiment. We then create dataset pairs from these clone injected datasets and label the dataset pair as either \textit{ClonePair} or not. Towards this end, we first collect tabular datasets that are commonly used in machine learning applications. Then we inject Type-1 and Type-3 data clones into them. Finally, we create the dataset pairs and label them. We explain each step in detail below. 

\noindent\textbf{Step 1: Dataset collection.} We select the tabular datasets from the UCI Machine Learning dataset repository~\cite{Dua:2019} for creating our synthetic dataset. This repository contains 439 tabular machine learning datasets for various machine learning tasks (e.g., classification, regression, and clustering). It is commonly used in research as a source of benchmark datasets for evaluating and comparing different machine learning algorithms~\cite{yoon2018pategan,papamakarios2017masked,alabdulmohsin2022reduction}. For our experiment, we select 154 datasets out of the 439 datasets by using the following three criteria: 1) a dataset must have more than 5 rows, to ensure that the dataset is large enough to be useful for machine learning tasks. 2) a dataset must have more than 2 columns, to ensure that the dataset has enough features or attributes to be informative for machine learning tasks. 3) a dataset must be parseable by the Python library Pandas~\cite{reback2020pandas} to enable ease of experimentation. To reduce processing time and memory usage, we truncate the datasets to have a maximum of 4000 rows and 60 columns (mean number of rows and columns across the selected 154 datasets). Table~\ref{tab:dataset_stat} provides details about our selected datasets.

\begin{table}
\caption{Basic details about UCI repository}
\centering
\small
\label{tab:dataset_stat}
\begin{tabular}{m{4em}m{3em}m{3em}m{3em}m{4em}m{4em}} \hline
 Mean \#Row & Mean \#Col & Max Row & Max Col & Total datasets & Selected datasets \\ \hline
1,774.8 & 24.92 & 4,000 & 60 & 439 & 154 \\ \hline
\end{tabular}
\end{table}

\noindent\textbf{Step 2: \textit{ClonePair} generation.} This step is comprised of three sub-steps. First, we create dataset pairs and then we inject them with either Type-1 or Type-3 clones depending on the type of dataset pair that we create. Finally, we label the dataset pairs as either \textit{ClonePair} or not. Algorithm~\ref{alg:injection} shows our overall approach for injecting data clones and labeling \textit{ClonePairs}.

\noindent\textbf{Step 2.1: Dataset pair creation.} We take the 154 datasets, and create pairs among them as shown in lines 1 and 2 of the Algorithm~\ref{alg:injection}. This creates two types of dataset pairs. First, identical pairs where a dataset A is paired with itself and non-identical pairs where dataset A is paired with other datasets in the list. This step would create 154 identical pairs and 11,781 non-identical pairs. 

\noindent\textbf{Step 2.2: Clone injection}. We inject Type-1 clones on identical pairs (Lines 7-8) and Type-3 clones on the non-identical pairs (Lines 9-10) using the following operations.

%After acquiring the datasets, we generate data pairs, both pairs consider as clone and non-clone, as described as shown in Algorithm~\ref{alg:injection}, in which 
%We design two types of operations to simulate the two types of data clones, type 1 and type 3 as follows: 

\begin{figure}
    \centering
    \includegraphics[width=8cm]{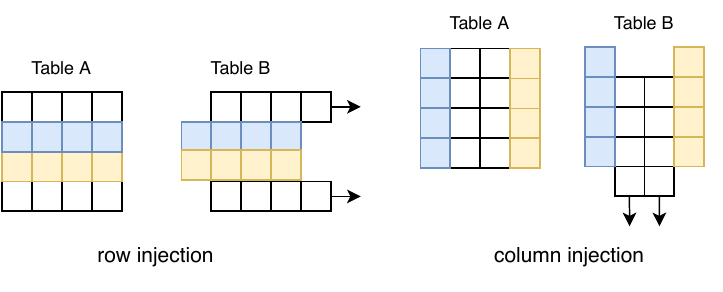}
    \caption{Type-1\textunderscore injection operation.}
    \vspace{-0.1in}
    \label{fig:type1}
\end{figure}

\begin{figure}
    \centering
    \includegraphics[width=8cm]{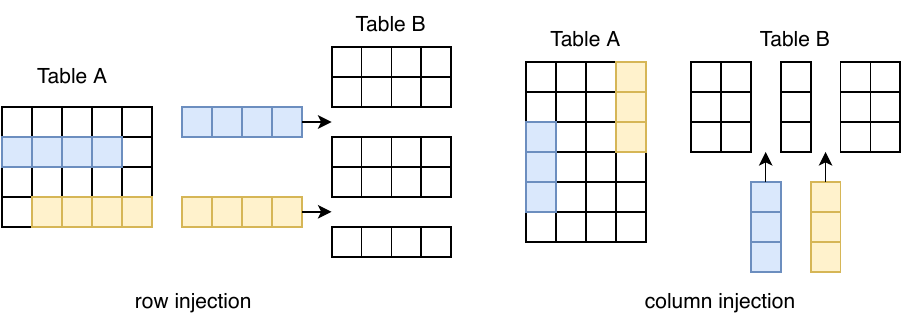}
    \caption{Type-3\textunderscore injection operation.}
    \vspace{-0.1in}
    \label{fig:type3}
\end{figure}

\begin{enumerate}[leftmargin=*]

\item \textbf{Type-1\textunderscore injection($A$, $p$):} Figure~\ref{fig:type1} presents an outline of Type-1\textunderscore injection operation. Given an identical pair (e.g., ($A$,$A$)) only one instance of the dataset $A$ is passed on to the operation. For the dataset $A$, we randomly sample $p$\% of all consecutive columns/rows (i.e., exact clone) from $A$ and create a new dataset $B$ using sampled columns/row. Therefore, we generate a data pair $A$ and $B$ with Type-1 clone. Note that columns and rows are randomly sampled with 50\% and 50\% probability. For instance, when setting $p = 20\%$, we randomly generate a percentage of 20\% of columns or rows (by random choice) from $A$ to create a new table $B$ with those sampled columns or rows.

\item \textbf{Type-3\textunderscore injection($A$, $B$, $p$):} Figure~\ref{fig:type3} presents an outline of Type-3\textunderscore injection operation. Given a non-identical dataset pair $A$ and $B$, we randomly sample $p$\% of all columns/rows from $A$ and inject them into $B$. Note that, we inject the sampled columns/rows in the random positions of $B$. Therefore, the order of sampled columns and rows may vary in $B$. In addition, the shapes of $A$ and $B$ may be different. To handle this, we always sample columns/rows from the bigger tabular dataset and inject them into the smaller dataset, by dropping the extra. Similar to Type-1\textunderscore injection operation, columns, and rows are randomly sampled with 50\% and 50\% probability.
\end{enumerate}

\noindent\textbf{Step 2.3: \textit{ClonePair} labeling.}
We label if the given dataset pair (both identical and non-identical pair) is a \textit{ClonePair} based on a threshold $t$. We label the pair having at least $t$ cloned data as a \textit{ClonePair} (Line 12 - 13) and otherwise label it as non-ClonePair (Line 14-15). In a real-world scenario, practitioners probably consider the different amounts of duplicated data as clones. Therefore, we keep $t$ as a threshold that allows practitioners to tune and train the clone detection classifier with different degrees of sensitivity. To reflect this in the algorithm, we first generate a random number from [0, t) and [t, 1] with equivalence chance (Line 3-5). 

%The algorithm generates Type 1 clone if the datasets are identical using $Operation1$ (Line 7-8), otherwise generates Type 3 clone using $Operation2$ (Line 9-10). In other words, we generate the Type 1 clone for every dataset in the list.

In summary, for a total of 154 datasets, we generated 11,935 dataset pairs. Algorithm~\ref{alg:injection} involves the random process, the number of \textit{clonePairs} and non-clone pairs may vary for different $t$. However, we still maintain an approximate ratio of 1:1 for non-ClonePairs and \textit{ClonePairs} because of the design of our algorithm. For instance, we generate 5552 \textit{ClonePairs} and 5474 non-ClonePairs, respectively, when $t = 10\%$.

\begin{algorithm}
\caption{Data clone injection}\label{alg:injection}
% \scriptsize
\KwIn{$table\_list$: A list of tabular datasets}
\KwIn{$t$: A threshold that is used to label data pairs}
\KwOut{$data\_pairs\_clone$: A list of clone data pairs}
\KwOut{$data\_pairs\_nonclone$: A list of non-clone data pairs}

\For{$i \gets$ $0$ to $size(table\_list)$ in $table\_list$}{
        \For{$j \gets$ $i$ to $size(table\_list)$ in $table\_list$}{
            \tcc{Randomly generate a number from $[0,t)$ or $[t,1]$ with equivalence chance}
            $p1$ = randomly generate a number $\in [0,t)$ 
            
            $p2$ = randomly generate a number $\in [t,1]$
            
            $p$ = randomly pick from $p1$ and $p2$ with equivalent chance
            
            $generatedTable$ = NULL
            
            \eIf{$i = j$}{
                
                $generatedTable$ = \textbf{Type-1\textunderscore injection}$(table\_list[i], p)$
            }{
                $generatedTable$ = \textbf{Type-3\textunderscore injection}$(table\_list[i], table\_list[j], p)$
                
            }
            \tcc{Label the data pair based on the amount of duplicated data.}
            \eIf{$p \ge t$}{
                $data\_pairs\_clone$.add($table\_list[i]$, $generatedTable$)
            }{
                $data\_pairs\_nonclone$.add($table\_list[i]$, $generatedTable$)
            }
        }
}
\end{algorithm}
% \begin{algorithm}
% \caption{data clone injection}\label{alg:injection}
% \end{algorithm}
%  \Input{$Set_{dataset}$}
%  \Output{$\langle D_i, D_j\rangle \in Set_{dataset}$}
%     \FOR{dataset $i$ in $Set_{dataset}$}
%         \FOR{dataset $j$ in $Set_{dataset}$}
%             \IF{$i = j$}
%                 \STATE simulate Type-1 Clone
%                 \STATE do injection operation 1
%             \ELSIF{$j > i$}
%                 \STATE simulate Type-3 Clone
%                 \STATE do injection operation 2
%             \ENDIF
%         \ENDFOR
%     \ENDFOR
% \end{algorithm}
% \caption{data clone injection}\label{alg:injection}
% \begin{algorithmic}
% \STATE $t \gets threshold$
% \STATE \sw{how did you select two datasets? are you using a nested loop to go through each pair, then see if it is a self-pair or not?}
% \IF {table pair is self-pair \sw{what is self-pair}}
%   \STATE do type-1 data clone injection $\theta 1$
%     \STATE 1/2 chance do row slicing from 0\%-t\%
%     \STATE 1/2 chance do column slicing from 0\%-t\%
% \ELSE 
%     \STATE 1/2 chance do type-3 data clone injection $\theta 2$
%   \IF {do type-3 data clone injection}
%     \IF { two tables have the same row structure }
%     \STATE do row injection from t\%-100\%
%     \ELSE
%     \STATE do column injection from t\%-100\%
%     \ENDIF
%    \ELSE
%    \IF { two tables have the same row structure }
%     \STATE do row injection from 0\%-t\%
%     \ELSE
%     \STATE do column injection from 0\%-t\%
%     \ENDIF
%   \ENDIF
% \ENDIF 
% \end{algorithmic}

% \sw{update based on the description. table $A$ and $B$. clone injection to column inject. remove the white columns in table 1*, since we create a new one using the sampled data, rather than injecting into an existing one.}

%In order to train our classifier, we require features that will help our classifier capture the presence of data clones in a dataset pair. In our study, we use value-based similarity between columns and rows as the features to train our classifier to identify data clones. 
%\subsection{Dataset-pair creation}\label{sec:create-pair}
%In the training phase, we take ${n}$ datasets and generate all possible dataset pairs of the ${n}$ datasets. This process ensures that each dataset is paired with every other dataset in the set, resulting in a total of ${n \choose 2}$ dataset pairs along with their corresponding label (i.e., whether data clones exist in the pair). 
%Subsequently, we use pandas library in python to read and parse the datasets, and convert into the Pandas DataFrame format. Our task is to conduct data clone detection at the granularity of the dataset pair level, with the goal of identifying the presence of any data clones between the dataset pairs.

\subsection{Similarity computation}\label{sec:feature}
%In order to train our classifier, we require features that will help our classifier determine if a dataset pair is a \textit{ClonePair}. 
Prior techniques (e.g., LTC~\cite{zhang2020learning} and TableCheck~\cite{Dou2016}) that are designed for table clone detection on spreadsheets do not work effectively on tabular datasets, which usually are stored in formats like CSV which do not contain formatting attributes such as cell formula, font size, and font color.
Therefore, we need to design features that are value-oriented and can capture the similarity between two columns or two rows based on their cell values. Towards this end, we design various metrics to measure the similarity of two lists of values for different data types (i.e., numeric and string). Before diving into the calculation of features, we first introduce the similarity metrics we used in SimClone below. 

\subsubsection{Value-based similarity metrics}
%\sw{R2.5}
Measuring the similarity of two rows or columns is essentially measuring the similarity of two lists of cell values from various aspects. Tabular datasets could have two types of values, namely numeric and string. In general, we have two criteria when selecting metrics: 1) computationally efficient; and 2) effective in capturing the similarity between strings or numbers.
For string type, metrics are selected to capture the text similarity between two lists of strings. We assume that two clone lists of strings should have a portion of common words and share similar semantics. Therefore, we first select the following metrics, Jarcard, Levenshtein, and Simhash by following previous studies~\cite{jaccard,simhash}. Simhash is an efficient approach to estimating how similar two sets are by projecting the string into a hashing code~\cite{charikar2002similarity}. Two strings sharing the same hashing code are considered similar. Levenshtein is a traditional way to measure two strings' similarity~\cite{Levenshtein}, which is used to measure the minimum efforts of converting one string to another from the character level. 
Jaccard is widely used to measure the word overlap of two strings~\cite{jaccard}. However, all the above three metrics are sensitive to long strings. In our case, the size of data clone is uncertain (which could be long). To complement the above metrics, we select TextRank to identify the important words from two lists first, then calculate their similarity based on the overlap in the important words.
For numeric type, we assume that two clone numeric lists should have the same/similar distribution. Therefore, we calculate the similarity of two lists of numeric values by measuring the similarity of mean and deviation between two lists of numeric values, i.e., Mean and Deviation Similarity. We select mean and deviation since they are computationally efficient and are widely used to compare two distributions in statistical analysis~\cite{kobayashi2000comparing,barde2012use}. We also select Jaccard Similarity to measure the overlap between two numeric sets as complementary to capture the detailed element overlap. Combining multiple similarity metrics enables us to comprehensively measure the similarity between data which can in turn lead to more accurate results. We present details about the six similarity metrics that we use in SimClone in Table~\ref{tab:similarity_metrics}.

 %Note that Jaccard similarity works both for string and numeric types. Combining multiple similarity metrics enables us to comprehensively measure the similarity between data which can in turn lead to more accurate results. We present details about the six similarity metrics that we use in SimClone in Table~\ref{tab:sim_metric}.

Even though we use the similarity metrics given in Table~\ref{tab:similarity_metrics} in this paper, SimClone is inherently customizable. One can choose to incorporate other similarity metrics depending on their specific needs. It is worth noting that similarity measuring on string and numeric types can be a complex task and algorithms with higher accuracy often require more computational resources. As such, we have balanced the trade-off among training, inference time, and accuracy by limiting our selection to string-based~\cite{Farouk_2019} and hashing-based methods. %Other text similarity measures (e.g., Tf-Idf and Word2Vec~\cite{word2vec}) and distribution similairty measures (e.g., Kullback-Leibler divergence~\cite{joyce2011kullback}) may be considered if higher accuracy is desired.

\begin{table}[htbp]
  \centering
  \caption{Similarity metrics used in SimClone}
  \renewcommand{\arraystretch}{1.5}
  \large % Reduce font size for better fit
  \begin{tabular}{p{1.6cm}|p{1.5cm}|p{18em}|p{13em}}
    \hline
    \textbf{Similarity} & \textbf{Applicable Data Type} & \textbf{Formula} & \textbf{Description} \\
    \hline
    Jaccard & numeric, string & $Jaccard_{Sim}(S_{1}, S_{2}) = \frac{|S_{1}\cap S_{2}|}{|S_{1}\cup S_{2}|}$ & Measures similarity by comparing common/unique elements of two sets. \\
    \hline
    Simhash & string & $Simhash_{Sim}(S_{1},S_{2}) = \frac{Hamming\_Dist(Simhash(S_{1}),Simhash(S_{2}))}{64}$ & Hashes the sets into 64-bit sequences and compares using hamming distance. \\
    \hline
    Levenshtein & string & $Levenshtein_{Sim}(S_{1},S_{2}) = \frac{Levenshtein Distance(S_{1},S_{2})}{|S_{1}|+|S_{2}|}$ & Measures similarity as minimum efforts to convert one string to another. \\
    \hline
    TextRank & string & $TextRank_{Sim}(S_{1},S_{2}) = \frac{|S_{1}\cap S_{2}|}{\log |S_{1}|+\log |S_{2}|}$ & Similar to Jaccard but less sensitive due to the log operation. \\
    \hline
    Mean & numeric & $Mean_{Sim}(S_{1},S_{2}) = 1-\frac{abs(mean(S_{1})-mean(S_{2}))}{abs(mean(S_{1})+mean(S_{2}))}$ & Calculates similarity between the mean values of the sample sets. \\
    \hline
    SD & numeric & $SD_{Sim}(S_{1},S_{2}) = 1-\frac{abs(SD(S_{1})-SD(S_{2}))}{abs(SD(S_{1})+SD(S_{2}))}$ & Calculates similarity between the SD values of the sample sets. \\
    \hline
  \end{tabular}
  \label{tab:similarity_metrics}
\end{table}

\subsubsection{Similarity matrices calculation}\label{sec:simmat}
Our aim in this step is to calculate the similarity values between the dataset pairs that we created in Section~\ref{sec:exp_data} to train data clone detection classifier. We choose to calculate the similarity at the granularity of the row and column level (i.e., calculate the similarity between each pair of rows/columns in the paired datasets) instead of at the cell level (i.e., where each cell in a dataset is compared to all other cells of the other dataset) or tabular level (i.e., the two tables are compared against each other based on the table properties) for the following reasons: 1) Calculating similarity at the tabular level might be too coarse-grained. Also, at the tabular level, comparisons are typically unable to capture row/column characteristics. For example, suppose only a small portion of data is cloned in two tables, while the majority of data is different from each other. The contribution of the small portion of cloned data might not be captured by the feature. 2) Calculating the similarity at the cell level is extremely time and space-consuming. For example, calculating the similarity for any similarity metric of two datasets with a size of 4000 rows x20 columns at the cell level would result in a matrix of size ($4000^{2}\times 20^{2}$), requiring $2^{11}$bits of space when using float32. Therefore, to optimize the time and space complexity, we use the row/column level of granularity to calculate the similarity.

Before calculating the similarity between a dataset pair, we divide each dataset $D$ in the dataset pair into two sub-datasets: $D_{string}$ and $D_{numeric}$, according to the data type of columns. For instance, suppose we have a dataset with 5 columns of string type and 3 columns of numeric type. We will divide the dataset into two sub-datasets, one with 5 columns of string type and another with 3 columns of numeric type. So each dataset pair will have 4 sub-datasets (two sub-datasets per dataset in a dataset pair). We perform the splitting to decrease the size of the generated similarity matrix and optimize calculation time subsequently. 
%This is done as the aforementioned similarity measures (with the exception of Jaccard Similarity) are both applicable for strings or numbers. 

\begin{figure}
    \centering
    \includegraphics[width=8cm]{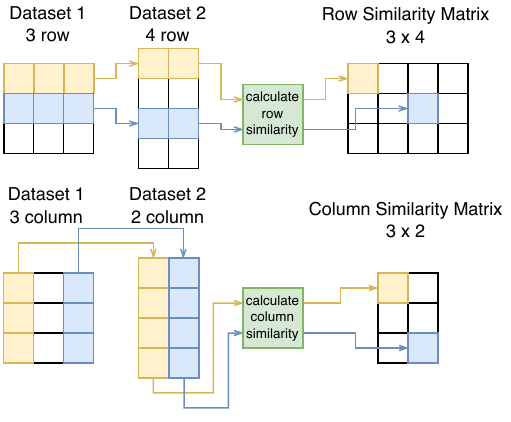}
    \caption{Calculation of row-to-row \& column-to-column similarity matrices.}
    \vspace{-0.1in}
    \label{fig:rowsim}
\end{figure}

We show how we calculate the similarity between two sub-datasets at the row/column level in Figure~\ref{fig:rowsim}. All of the value-based similarities mentioned above are calculated at the row-row/column-column level. We can express the calculation of the similarity matrix as:
$$ M_{SimVF_x}^{row}{i,j} = Sim{VF_{x}}(D1_{datatype}^{row_{i}},D2_{datatype}^{row_{j}})$$
$$ M_{SimVF_x}^{col}{i,j} = Sim{VF_{x}}(D1_{datatype}^{col_{i}},D2_{datatype}^{col_{j}})$$
where $SimVF_{x}$ denotes the value-based similarity metrics presented in Table~\ref{tab:similarity_metrics}. $M_{SimFV_x}^{row}$ denotes the resultant matrix for row-row level similarity using $SimVF_x$ between sub dataset $D1_{datatype}$ and $D2_{datatype}$ and $M_{SimVF_x}^{col}$ denotes the resultant matrix for column-column similarity. $Datatype \in \{numeric, string\}$. $D1_{datatype}^{row_{i}}$ denotes the row $i$ in $D1_{datatype}$ and $D1_{datatype}^{col{j}}$ denotes the column $j$ in $D1_{datatype}$. It is worth noting that the shape of the resultant similarity matrix is decided by the number of rows/columns in the dataset pair. For instance, two datasets with 10 and 20 rows respectively lead to a row-row matrix with a size of 10 x 20.
In our approach, we have four similarity metrics for string types, and three similarity metrics for numeric types. We calculate each similarity metric in row-row level ($ M^{row}$) and column-column level ($ M^{col}$) according to the data type. Therefore, we obtain 14 similarity matrices between each paired dataset, i.e., 7 similarity matrics for row-row level and another 7 for column-column level. Note that we do not have to calculate similarity for the missing data type and we pad the missing ones with zero.

\begin{figure*}[ht]	
    \includegraphics[width=\textwidth]{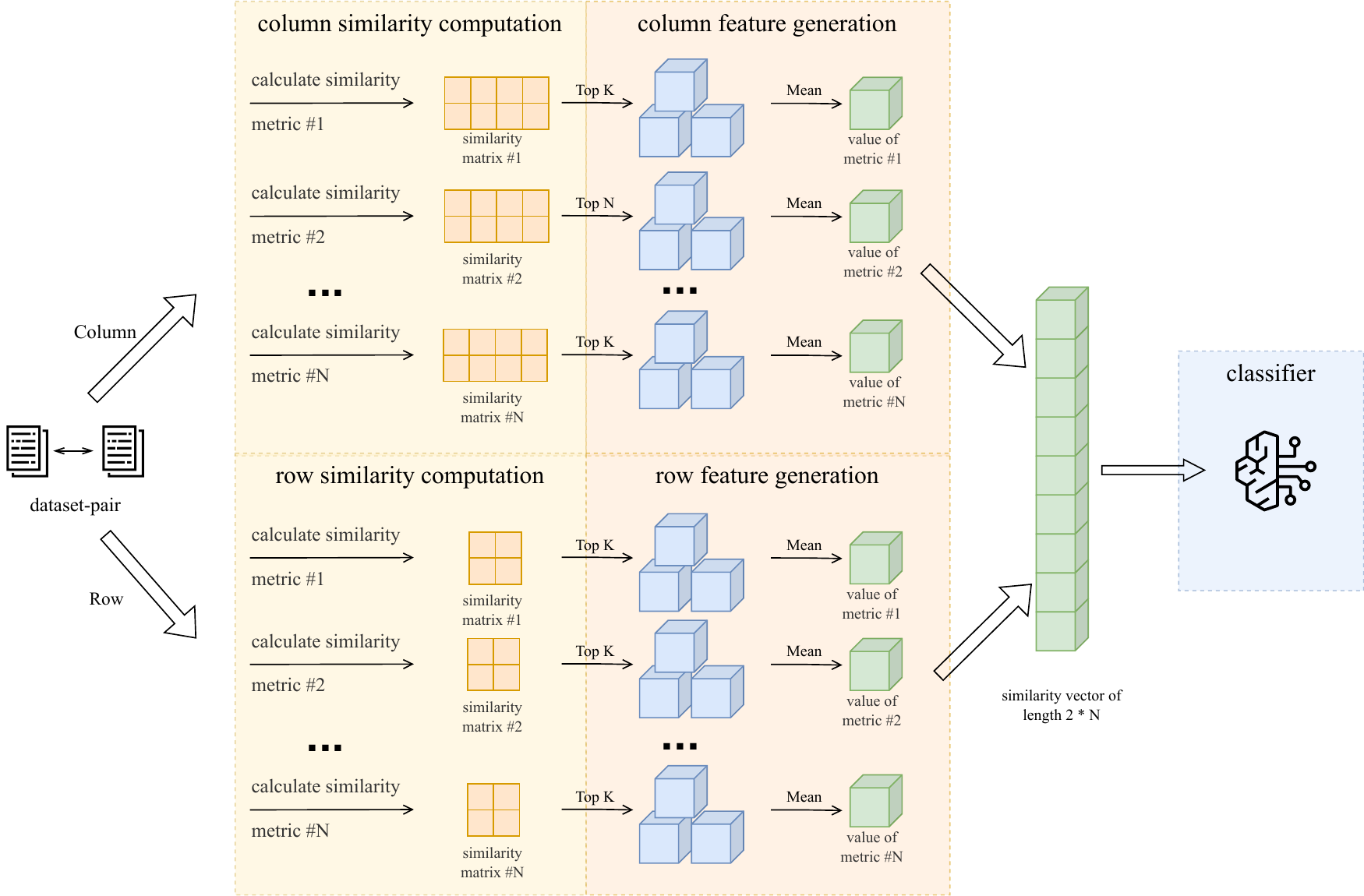}
    \caption{Workflow of similarity computation and feature generation for a dataset-pair. }
    \label{fig:pipeline}
\end{figure*}

\subsection{Feature generation}

In this stage, we utilize pooling technique to process the similarity matrices generated in the previous step~\ref{sec:simmat} and convert them into features that are used by a machine learning classifier. The pooling technique extracts relevant information from a system's output and converts it into a uniform, usable format (e.g., consistently picking the top 10 values in a matrix and flattening a matrix). This concept is widely used in both neural networks and other domains like computer vision and graph theory~\cite{zhou2020graph}.
We use pooling technique since the size of the resultant similarity matrices varies among dataset pairs with different sizes of columns and rows as shown in Section~\ref{sec:simmat}. We leverage the pooling technique to obtain a uniform output, with a fixed size, that can be used as the input for the classifier in the following step. We do so to ensure that we can process all dataset pairs in a consistent format.

In SimClone, we design a pooling calculation called ``Mean Top K'', which extracts the K largest values from a given similarity matrix $M$ and calculates their mean value as the pooling value. Each value in the similarity matrix represents the similarity between two rows (or columns), and most of the values in the similarity matrix are very small. Therefore we hypothesize that large similarity values are more likely to signal the presence of clones and we only need to pay more attention to the part with larger values. We calculate the pooling value as given below:
$$pooling\ value = \frac{\sum(\operatorname{argmax}_{m^{\prime} \subset M,\left|m^{\prime}\right|=K})}{K}$$

This pooling calculation is applied to each of the 14 similarity matrices generated in Section~\ref{sec:simmat}, resulting in a total of 14 values which are then concatenated into a single vector of size to serve as the similarity feature vector for a given dataset pair. We set the value of K to be 10 in our SimClone method. This feature vector can then be used as input for the classifier. Note that if a similarity matrix is missing from the previous step (i.e., when only one type of data is present in a dataset pair), we simply generate a pooling value of 0 for it.

Figure~\ref{fig:pipeline} presents the detailed workflow of similarity computation and feature generation. Given a dataset-pair, as the flow shows, we first calculate all the similarity metrics at the row and column levels and produce a similarity matrix for each similarity metric. Then in the feature generation, we perform Mean Top K pooling to get a mean value for each similarity matrix as the representative for the corresponding similarity matrix. Finally, we concatenate all the mean values into a single vector as the input for the classifier. 

\subsection{Data clone detection classifier construction and inference}\label{sec:classifier}

After obtaining the unified feature vector representation for all the dataset pairs, we construct a binary classifier to predict if a dataset pair is a \textit{ClonePair}. We train our classifier using a dataset of labeled dataset pairs to predict the likelihood of a given dataset pair being \textit{ClonePair}. After pooling, the feature can be used to train all kinds of machine learning classifiers, in this study, we use Random Forest (RF)~\cite{biau2016random}, XGBoost~\cite{chen2016xgboost}, CatBoost~\cite{prokhorenkova2018catboost}, and LightBGM~\cite{ke2017lightgbm} as our classifiers, since they are widely used and achieved competitive performance for classification tasks~\cite{zhang2020learning,szczepanek2022daily,bentejac2021comparative}. %In the section\ref{sec:exp}, we will provide more information on the training, testing, and inferencing process for this classifier.

In the inference phase, given a new collection of datasets, we create data pairs, calculate similarity metrics, and generate features as described above. We then use our trained classifiers to predict if any given dataset pair is a \textit{ClonePair}.

\subsection{Data clone visualization} \label{sec:vis}
\begin{figure}[]
	\includegraphics[width=\columnwidth]{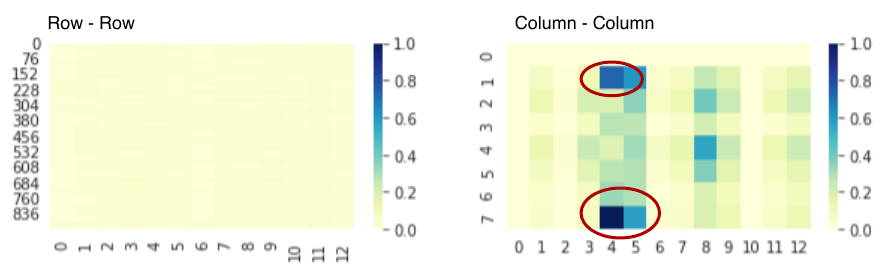}
    \caption{Example of SHAP-based visualization. Darker part indicates Clone (in red circle).}
    \label{fig:shapviz}
\end{figure}

In this step, we describe how we create our visualization approach to help the users of SimClone identify the exact location of data clone in a \textit{ClonePair}. Such a visualization is particularly useful when the datasets in the \textit{ClonePair} contain a significant number of data points. We present an overview of our visualization approach in the visualization phase part of Figure~\ref{fig:pipeline}. 

%\sw{I try to explain the rationale behind using var importance, double check on it}
We combine the feature importance scores that we obtain by interpreting classifier that we construct in the previous step with the similarity matrices obtained in Section~\ref{sec:simmat}. The intuition behind combining similarity matrices with the feature importance scores is as follows: 
the similarity matrix by itself is a good indicator of clone location in the \textit{ClonePair}. For instance, a large similarity value in the matrix $M^{row}{i,j}$ indicates that rows $i$ and $j$ are similar to each other and might indicate the presence of a clone. However, we have 14 similarity matrices, and returning all of them to the end-users could cause confusion and information overload. Moreover, all similarity metrics are not equally important. One similarity matrix may be a better indicator than another at characterizing the presence of a clone. Therefore, we weight the similarity matrices with the feature importance scores obtained from the classifier trained in the previous step to amplify the contribution of the important similarity metrics that help the classifier identify the \textit{ClonePair} and vice versa. 

%we first need to identify the importance of each similarity matrix. Recall that each feature is readout from a similairty matrix. We need to select a local classifier explanation technique to inform us which features are more important for each prediction. In this study, 
In our SimClone method, we utilize the SHapley Additive exPlanations (SHAP)~\cite{shap} to compute the feature importance scores at the instance level (i.e., each dataset pair). We use SHAP since it is widely used by prior studies to compute the local feature importance scores~\cite{XuICSE2023,rajbahadur2021impact,esteves2020understanding}. SHAP calculates the Shapley value for each feature and uses the Shapley value as the proxy to estimate the contribution or the importance of each feature towards the classifier's prediction. Next, we take SHAP values for each value-based similarity feature and perform a dot product with the corresponding similarity matrix obtained for each dataset pair. This step allows us to assign different weights for similarity matrices according to their importance to the classifier's prediction. Since the shapes of the row-row matrix and column-column matrix are different, we cannot merge them into one matrix. Therefore, after the weighted combination, we end up with two matrices $M_{vis}^{row}$ and $M_{vis}^{col}$, and they are calculated as follows:

$$M_{vis}^{row} = \frac{1}{n}\sum_{i=1}^{n} (SHAP_i \cdot M_{SimVF_i}^{row})$$
$$M_{vis}^{col} = \frac{1}{n}\sum_{i=1}^{n} (SHAP_i \cdot M_{SimVF_i}^{col})$$
, where $n$ is the number of similarity metrics (i.e., 7), $M_{SimVF_i}^{row}$ and $M_{SimVF_i}^{col}$ are the resultant similarity matrices generated in Section~\ref{sec:simmat}, $SHAP_i$ is the importance score for each value-based similarity feature.

Finally, we plot the weighted similarity matrices as heat maps, with deeper colors indicating a higher likelihood of the presence of a clone (i.e. red circle in Figure~\ref{fig:shapviz}). This allows for clear and intuitive visualization of the exact location of data clones found in a \textit{ClonePair}.

\section{Experiment Design}\label{sec:exp}
In this section, we first introduce the research questions. Then, we explain how we collect the datasets, build our baselines, and how we answer each research question in detail. 

%In this Section, we first outline how we synthetically create a data clone injected dataset to train our SimClone method and explain our training process. 
%\sw{formulated the structure in RQ style, please check} \sw{for rq1-rq3, we use $t$=10\%, and rq4, we investigate different values of $t$.}
\subsection{Research Questions}
We formulate the following research questions to evaluate SimClone from various aspects: 

\begin{itemize}
    \item RQ1: How effective is SimClone in identifying \textit{ClonePairs} compared to the state-of-the-art (SOTA) baselines?
    %\item RQ2: How effective is SimClone in identifying \textit{ClonePairs} compared to SOTA baselines on real-world datasets?
    \item RQ2: How effective is SimClone's visualization approach? 
    \item RQ3: Which similarity metrics make the most contributions towards the effectiveness of SimClone?
    \item RQ4: How does SimClone's effectiveness change with different threshold $t$?
\end{itemize}

In RQ1, we evaluate how effectively our SimClone method is capable of identifying \textit{ClonePairs} compared to the SOTA tabular clone detection method LTC (i.e., Detection performance evaluation) on synthetic and real-world dataset, respectively. 
In RQ2, we evaluate how effective the visualizations generated by our SimClone method are at locating the exact location of data clones in \textit{ClonePair} (i.e., Visualization performance evaluation). SimClone uses six different similarity metrics and could be time-consuming, in RQ3, we investigate the contribution of each similarity metric and investigate the possibility of building a lightweight version of SimClone. In RQ4, we investigate how SimClone's effectiveness changes with various threshold $t$. 

\subsection{Baseline for identifying \textit{CloneParis}}
As far as we are aware, we are the first to propose a data clone detection method specifically for tabular datasets used in machine learning. Hence, we have chosen to use table clone detection methods, which were originally intended for data clone detection in spreadsheets, as our baseline. We do so as they are the most comparable method to our method. The LTC approach~\cite{zhang2020learning} is currently considered SOTA in table clone detection in spreadsheets. However, LTC was developed for spreadsheets, and it leverages formatting features like font, and cell color. We do not have such information in our dataset. To adopt LTC for our dataset we drop the format features like font size, font color (which LTC method uses), and only use the column header, row header, and cell type features to build LTC. Since the code for LTC is not publicly available, we implemented our own version in Python based on their paper. %\sw{To validate whether our implementation is correct or not, we evaluate our implemention on ? dataset, and we found that the performance is ?, which is smilar to the results reported on their paper.}\xy{Not feasible}

\subsection{Data preparation and evaluation}\label{sec:dataPrep}
To evaluate SimClone's effectiveness in identifying \textit{ClonePairs}, we prepare two datasets, the synthetic dataset and the real-world dataset. We elaborate on how we construct the datasets and evaluate them below. 

\subsubsection{Synthetic dataset}
we first evaluate SimClone on the Synthetic dataset that we constructed based on UCI repository (as we introduce in Section~\ref{sec:exp_data}) and compare it to the baseline. 

We compute five commonly used classification evaluation metrics: accuracy, F1-score, precision, recall, and Area Under the Curve (AUC) for both methods on all the folds. We chose these metrics since they are commonly used by prior studies to evaluate a classifier's performance~\cite{zhang2020learning,rajbahadur2021impact,XuICSE2023}. We report the average of the performance scores for each of the studied evaluation metrics for all the synthetic datasets generated for the studied thresholds across the 10 folds. 

\subsubsection{Real-world dataset}
To complement our evaluation of the synthetic dataset, we also conduct experiments on real-world spreadsheets that we obtained from the EUSES Spreadsheet corpus~\cite{Fisher2005} and Enron corpus~\cite{hermans2015enron}. EUSES corpus contains spreadsheets collected from the internet pertaining to various domains~\cite{Fisher2005} and is made up of over 4,000 real-world spreadsheets.
The Enron corpus was collected from the Enron email archive within the Enron corpus, and contains more than 15,000 spreadsheets. 
We explain the process of creating the real-world evaluation datasets below.

\begin{table}
\caption{Basic details about EUSES and Enron dataset}
\centering
\large
\label{tab:EUSES_stat}

\begin{tabular}{lllllll} \hline
Dataset &Mean \#Row & Mean \#Col & Max Row & Max Col & Total datasets & Selected \ spreadsheets \\ \hline

\multicolumn{1}{r}{EUSES}&\multicolumn{1}{r}{41.30} & \multicolumn{1}{r}{5.85} & \multicolumn{1}{r}{10,979} & \multicolumn{1}{r}{244} & \multicolumn{1}{r}{17,871} & \multicolumn{1}{r}{1,182} \\ 
\multicolumn{1}{r}{Enron}&\multicolumn{1}{r}{145.81} & \multicolumn{1}{r}{15.06} & \multicolumn{1}{r}{65,535} & \multicolumn{1}{r}{256} & \multicolumn{1}{r}{67,725} & \multicolumn{1}{r}{1,000} \\ \hline
\end{tabular}%

\end{table}

\noindent\textit{Step 1: Real-world dataset collection.} Several prior studies have noted that tables contained in the spreadsheets available in the EUSES and Enron corpus have clones and have used it in their study to evaluate their data clone detection methods~\cite{Dou2016,Hermans2013,xu2018detecting}. Therefore, we use the tables contained in the spreadsheets available in the corpus to evaluate SimClone's detection performance and compare it with LTC's detection performance. 

Both corpus contains spreadsheets that can contain multiple tables in one spreadsheet. For our SimClone method, each dataset is comprised of only one table. So we first parse the spreadsheets available in the corpus to extract the tables and create our real-world dataset. We do so through a Java tool that we developed using the Apache POI library\footnote{https://poi.apache.org/} to extract the tables from the spreadsheets. Our tool processed the corpus and generated 17871 tabular datasets for EUSES and 67725 tabular datasets for Enron. Then we select datasets by following three criteria: 1) a table must have more than 2 rows and 2 columns. 2) we only keep the table with a unique set of column headers to avoid the obvious influx of Type-1 clone. It is important to note that this process resulted in the loss of formatting information, such as font size and color, in the EUSES tables in csv format compared to their original Excel format. We ended up with 1,182 datasets for EUSES and 8,184 datasets for Enron. We further randomly sampled 1,000 datasets among the 8,184 datasets for Enron since using full datasets requires too many computation resources. Table\ref{tab:EUSES_stat} provides some basic details about the EUSES dataset.

\noindent\textit{Step 2: Real-world dataset pair generation.} We start by creating dataset pairs for the tabular datasets. We end up with a total of 69.8k dataset pairs for EUSES and 49.95k dataset pairs for Enron. Note that we do not inject any data clone into those dataset pairs since we aim to detect if SimClone can detect any data clones in real-world data.

\noindent\textit{Step 3: Detection performance evaluation.} Note that we do not have labels for real-world dataset pairs\footnote{Prior studies that use EUSES and Enron for clone detection do not make their labels public.} and it is impossible to label all pairs. Therefore, we use both SimClone and LTC that are trained from the synthetic dataset on all dataset pairs to identify the \textit{ClonePairs}. We then collect the top 200 \textit{ClonePairs} returned by both SimClone and LTC (sorted by the probability score provided by SimClone and LTC) for each corpus. Finally, we check every pair manually (400 pairs in total, 200 \textit{ClonePairs} returned by each method) and check if it indeed is a \textit{ClonePair}. Note that we consider a dataset pair to be a \textit{ClonePair} if the two datasets have at least one identical column or one row of data. Two first authors manually label the returned 400 \textit{ClonePairs} independently to evaluate if they were indeed \textit{ClonePairs}. After that, the two authors discussed the results to resolve any disagreements until a consensus was reached. The labeling process has a Cohen’s kappa of 0.89 for LTC and 0.72 for SimClone, which indicates a substantial level of the inter-rater agreement~\cite{viera2005understanding}. 

We calculate the Precision@K (short for P@K) with different K (i.e., 5, 10, 20, 50, 100, 200) to evaluate the effectiveness of SimClone and LTC on their respectively labeled \textit{ClonePairs}. We calculate Precision@K using $Precision@K = \frac{True\ clone\ pairs}{Top\ k\ returned pairs}$.

\subsection{Approach for RQ1: Effectiveness of SimClone}

We evaluate the effectiveness of SimClone on the synthetic and compared its performance with LTC using the metrics introduced in Section~\ref{sec:dataPrep}
Note that in this RQ, we set the threshold $t$ to 10\%. In other words, we train SimClone and LTC methods on the synthetic dataset using a threshold of 10\% and test them on both Synthetic and real-world datasets. We do this, since we want to use our SimClone method trained on the synthetic dataset where the \textit{ClonePairs} are labeled based on what we assume might be the most realistic threshold. We evaluate SimClone on the synthetic dataset with all studied classifiers.

In addition, we evaluate the effectiveness of SimClone on real-world datasets and compare its performance with LTC using the metrics introduced in Section~\ref{sec:dataPrep}. We only evaluate the performance of SimClone on Random Forest, since it performs the best among all studied classifiers. Our evaluation on the real-world dataset requires manual labeling, and labeling on all classifiers is not feasible.

\subsection{Approach for RQ2: Effectiveness of SimClone's Visualization Approach}

%\noindent\textbf{Baseline}
In Section~\ref{sec:vis}, we explained that the Similarity matrices by themselves might be a good indicator of the location of the data clone in a \textit{ClonePair}. As far as we know, our visualization approach is the first work on this, therefore, we construct a baseline by combining all the similarity matrices using equivalent weights instead of weighing them by the feature importance scores computed from the classifier for a given dataset pair as our baseline.

To evaluate the effectiveness of our SimClone method's visualization results, we generate a heatmap using our SimClone method on all true positive pairs (i.e., the pairs that are correctly identified as \textit{ClonePair}) identified by our SimClone method on one of the 10 folds that we randomly chose. The darker the color of an area in the heatmap, the higher the likelihood of that area being a clone. To quantitatively measure the accuracy of our proposed approach, we use popular metrics precision@K (P@K), where $K$ we set as 1, 5, and 10. Precision@K has been widely used in similar software engineering localization tasks, such as bug/fault localization~\cite{smytzek2022sflkit,rahman2018improving} and vulnerability detection~\cite{yang2023does,li2021vulnerability}. Our classifier returns \textit{ClonePairs} with likelihood, which indicates the classifier's confidence in the prediction results. We examine the effectiveness of our visualization approach on the predictions with different confidences. More specifically, we examine the top 20, 50, 100, and all prediction results by the classifiers. We compare the results provided by SimClone's visualization and those provided by the baseline. 

We do this evaluation on the synthetic dataset since we precisely know where the clones in a \textit{ClonePair} are located and we do not have such information about the real-world dataset. We use the threshold of 10\% and use the Random Forest as our classifier, since Random Forest performs the best among all studied classifiers.

\subsection{Approach for RQ3: Effectiveness of Different Similarity Metrics}
SimClone uses six different similarity metrics. Even though we calculate similarities at row and column level instead of cell level to reduce SimClone's time complexity, such calculation can still be expensive. For $n$ datasets, similarity metrics of ${n \choose 2}$ pairs need to be calculated. In previous sections, we examined the effectiveness of the metrics as a bundle. However, each metric may not contribute equally to the classifier. Therefore, identifying the most effective similarity metrics and reducing the number of similarity metrics can significantly improve the efficiency of SimClone. To identify the metrics that contribute the most towards the performance and cost the least amount of time we performed ablation experiments on SimClone. In particular, we compared the performance of classifiers using each similarity metric alone. We also recorded the time spent on calculating similarity metrics. Same as RQ2, we use the SimClone that we trained on the synthetic dataset with the threshold of 10\% to evaluate the performance of the visualization method and use the Random Forest as our classifier. 

\subsection{Approach for RQ4: Impact of Different Threshold $t$}
As we mention earlier in Section~\ref{sec:exp_data}, we label a dataset as \textit{ClonePair} or not based on a given threshold $t$. However, there is no universally accepted threshold for determining what percentage of similarity constitutes to make a dataset pair to be a \textit{ClonePair}. The threshold for determining clones can vary depending on the specific application or use case. For example, it is not clear whether the presence of three consecutive cells or whether ten percent of the data in a row being identical constitutes a dataset pair being a \textit{ClonePair}. These thresholds are typically defined by the user depending on the intended use of the data. Therefore, we aim to investigate the impact of threshold $t$ on the effectiveness of SimClone compared with the baseline LTC. We have designed a broad range of thresholds for our experiments, 0\%, 5\%, 10\%, 15\%, 20\%, 30\%, 50\%, 70\%, and 90\%. These thresholds will be used to determine whether cloned data between dataset pairs exceeds the threshold. If it does so, the given dataset pair is labeled as a \textit{ClonePair}. Therefore, we have 5 synthetic datasets, labeled based on the different thresholds upon which we train and test our SimClone method. For each of the synthetic datasets labeled using the aforementioned thresholds, we construct our SimClone method (and LTC method) using a 10-fold validation to ensure the robustness and reliability of our results. We use Random Forest as our classifier.

\section{Results} \label{sec:results}
\subsection{Results of RQ1: Effectiveness of SimClone}

\textbf{SimClone conclusively outperforms LTC at detecting \textit{ClonePairs} across all the studied metrics with a 0.851 F1-score and 0.923 AUC on the synthetic dataset.}
Table~\ref{tab:RQ1} presents the results of SimClone and LTC in terms of the studied evaluation metrics on the synthetic dataset. SimClone consistently outperforms LTC across all classifiers. For instance, SimClone achieves an AUC of 0.923 and F1-score of 0.851. When comparing the effectiveness of different classifiers, we notice that Random Forest performs the best compared to other classifiers across all studied metrics for SimClone, while for LTC, XGBoost performs the best. If we compare the best version of SimClone which uses RF with the best version of LTC which uses XGBoost, SimClone achieves an improvement of 26.1\% in terms of F1.

\begin{figure}[h!]
	\includegraphics[width=\columnwidth]{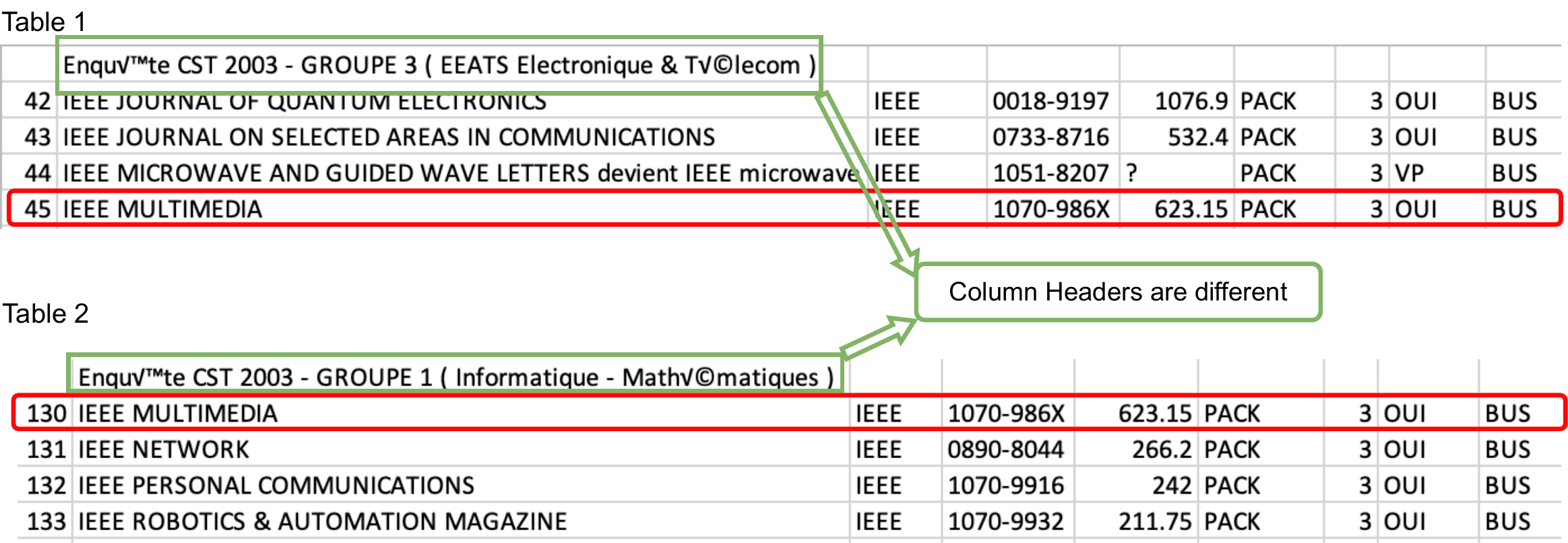}
    \caption{Example of a \textit{ClonePair} that SimClone correctly detects and LTC fails to identify. The red box indicates the data clone. The green box indicates minor changes to the column header. LTC fails to identify the clone since LTC relies on column header names while the column names in Table 1 and Table 2 are different.}
    \label{fig:rq1_example}
\end{figure}

%\subsection{Results of RQ2}
\textbf{SimClone outperforms LTC at identifying \textit{ClonePairs} on real-world dataset on both dataset in terms of Precision@K across all the studied values of K, except K = 10 on Enron.}
Table~\ref{tab:Euses} presents the results of SimClone and LTC on the real-world datasets (i.e., EUSES and Enron) by examining the top 200 results. On EUSES, SimClone consistently outperforms LTC across all values of K. For instance, We can see that Simclone still achieves a precision of 0.72, 0.57, and 0.44, even when examining the top 50, 100, and 200 returned pairs on real-world dataset in contrast to LTC, by achieving an improvement at least by 100\%. On Enron, we observe that in general SimClone outperforms LTC as well. However, the improvement margin is not as large as EUSES. Typically for Precision@10, LTC performs better than SimClone. The reason is that in Enron, a remarkable number of clone pairs returned by LTC share the same header name, therefore, LTC can leverage such information to identify clone pairs accurately, while SimClone does not leverage header information. However, it is worth noting that structural or formatting information such as row or column headers and formulas are not available in tabular datasets that are used in AI software development. The strength of SimClone method can be further reinforced by considering the example in Figure~\ref{fig:rq1_example}. LTC fails to detect the example as a \textit{ClonePair} since the column names are slightly different, however, since our method only relies on the value similarity, SimClone is able to correctly identify this example as a \textit{ClonePair}.

\begin{table}[]
\caption{The performance of SimClone and LTC on the synthetic dataset. We highlight the better-performed approach (SimClone V.S. LTC) with the same classifier with bold text.}\label{tab:RQ1}
\begin{tabular}{llrrrrrr}
\toprule
\textbf{Classifier} & \textbf{Approach} & \textbf{Accuracy} & \textbf{AUC}   & \textbf{F1}    & \textbf{Precision} & \textbf{Recall} \\
\midrule
\multirow{2}{*}{RF}      & SimClone & \textbf{0.850}    & \textbf{0.923} & \textbf{0.851} & \textbf{0.883}     & \textbf{0.820}  \\
                         & LTC      & 0.610    & 0.655 & 0.613 & 0.620     & 0.607  \\
\midrule
\multirow{2}{*}{XGBoost} & SimClone & \textbf{0.847}    & \textbf{0.920} & \textbf{0.845} & \textbf{0.873}     & \textbf{0.819}  \\
                         & LTC      & 0.645    & 0.686 & 0.674 & 0.632     & 0.722  \\
\midrule
\multirow{2}{*}{CatBoost}& SimClone & \textbf{0.850}    & \textbf{0.919} & \textbf{0.848} & \textbf{0.877}     & \textbf{0.822}  \\
                         & LTC      & 0.639    & 0.686 & 0.673 & 0.625     & 0.730  \\
\midrule
\multirow{2}{*}{LightBGM}& SimClone & \textbf{0.850}    & \textbf{0.923} & \textbf{0.848} & \textbf{0.872}     & \textbf{0.826}  \\
                         & LTC      & 0.636    & 0.694 & 0.669 & 0.624     & 0.722  \\
\bottomrule
\end{tabular}
\end{table}

\begin{table}
    \caption{The performance of SimClone and LTC when examining the top 200 returned results on EUSES and Enron real-world data in terms of Precision@K. We highlight the better-performed approach (SimClone V.S. LTC) with bold text.}\label{tab:Euses}%
% \resizebox{\columnwidth}{!}{%
    \begin{tabular}{cccccccc}
    \toprule
    
    \textbf{Dataset} & \textbf{Approach} & \textbf{P@5} & \textbf{P@10} & \textbf{P@20} & \textbf{P@50} & \textbf{P@100} & \textbf{P@200} \\
    \midrule
    \multirow{ 2}{*}{EUSES} & SimClone  & \multicolumn{1}{r}{\textbf{1.00}} & \multicolumn{1}{r}{\textbf{0.90}}   & \multicolumn{1}{r}{\textbf{0.85}}  & \multicolumn{1}{r}{\textbf{0.72}}  & \multicolumn{1}{r}{\textbf{0.57}} & \multicolumn{1}{r}{\textbf{0.44}} \\
    & LTC  & \multicolumn{1}{r}{0.40} & \multicolumn{1}{r}{0.40}  & \multicolumn{1}{r}{0.37}  & \multicolumn{1}{r}{0.36}  & \multicolumn{1}{r}{0.25} & \multicolumn{1}{r}{0.22} \\
     \midrule
   \multirow{ 2}{*}{Enron} & SimClone  & \multicolumn{1}{r}{\textbf{0.80}} & \multicolumn{1}{r}{0.60}   & \multicolumn{1}{r}{\textbf{0.70}}  & \multicolumn{1}{r}{\textbf{0.66}}  & \multicolumn{1}{r}{\textbf{0.66}} & \multicolumn{1}{r}{\textbf{0.44}} \\
    & LTC  & \multicolumn{1}{r}{0.60} & \multicolumn{1}{r}{\textbf{0.70}}  & \multicolumn{1}{r}{\textbf{0.70}}  & \multicolumn{1}{r}{0.52}  & \multicolumn{1}{r}{0.37} & \multicolumn{1}{r}{0.29} \\
    \bottomrule
    \end{tabular}%
    % }
  
\end{table}

\begin{comment}
\begin{table}
    \caption{The performance of SimClone and LTC when examining the top 200 returned results on Enron real-world data in terms of Precision@K.}\label{tab:Euses}%
% \resizebox{\columnwidth}{!}{%
    \begin{tabular}{ccccccc}
    \toprule
    
    \textbf{Approach} & \textbf{P@5} & \textbf{P@10} & \textbf{P@20} & \textbf{P@50} & \textbf{P@100} & \textbf{P@200} \\
    \midrule
    SimClone  & \multicolumn{1}{r}{0.80} & \multicolumn{1}{r}{0.60}   & \multicolumn{1}{r}{0.70}  & \multicolumn{1}{r}{0.66}  & \multicolumn{1}{r}{0.66} & \multicolumn{1}{r}{0.44} \\
    LTC  & \multicolumn{1}{r}{0.60} & \multicolumn{1}{r}{0.70}  & \multicolumn{1}{r}{0.70}  & \multicolumn{1}{r}{0.52}  & \multicolumn{1}{r}{0.37} & \multicolumn{1}{r}{0.29} \\
    
    \bottomrule
    \end{tabular}%
    % }
  
\end{table}
\end{comment}

\subsection{Results of RQ2: Effectiveness of SimClone's Visualization Approach}\label{sec:rq2}

% We evaluate if the highlighted area in the heap map truly contains a data clone. If the visualization correctly covers the data clone, we mark it as a ``hit''. 
% To evaluate the effectiveness of our visualization approach, we compare our approach with the \textit{baseline} that simply takes the mean value of all similarity matrices when grouping matrices into four categories without any weight assignment. 
% We consider the visualization as a recommendation system, and the heat map with the highest \sw{where is the average come from?}average heat value will be considered the result recommended by the system and the highest heat value in the heat map indicates the location of the data clone like the Figure~\ref{fig:shap_example} \sw{where is the figure? the you you use is the dot product example}.

\begin{figure*}[H]
    \centering
    \includegraphics[width=8cm]{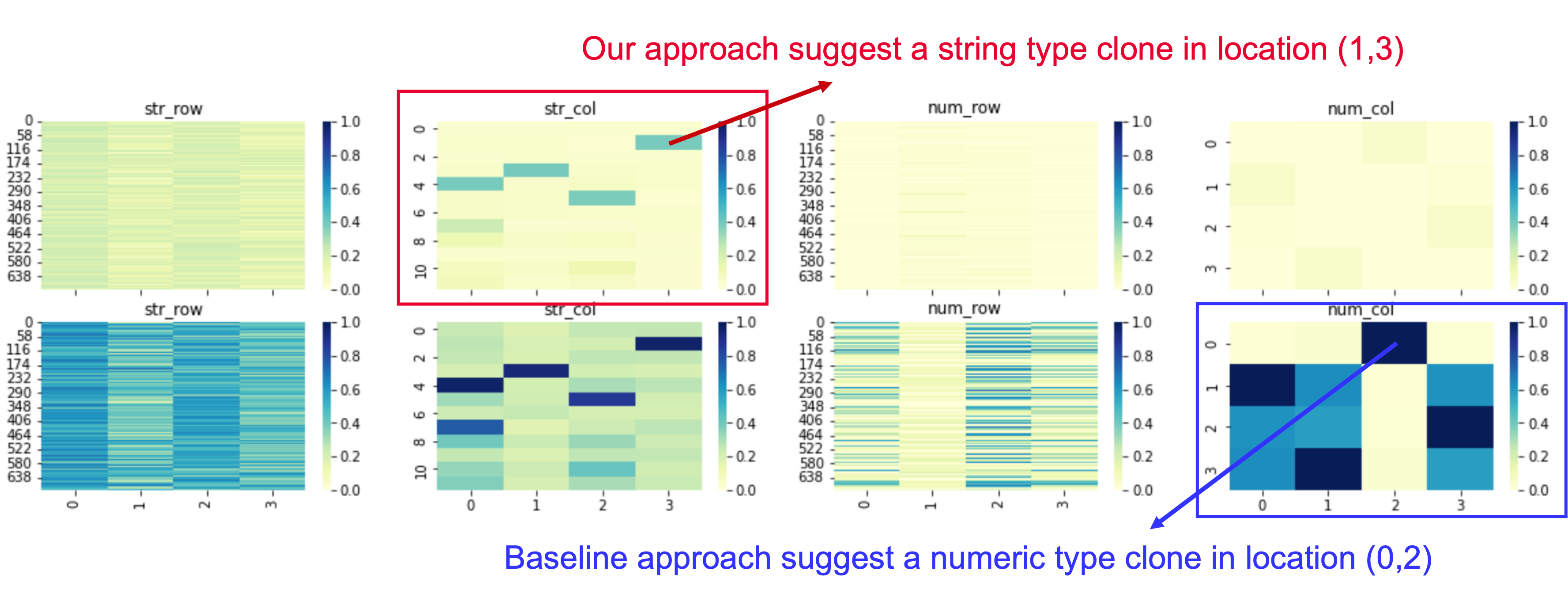}
    \caption{Different recommend result from our approach VS baseline}
    \label{fig:shap_demo}
\end{figure*}

\begin{figure}[]
	\includegraphics[width=0.5\columnwidth]{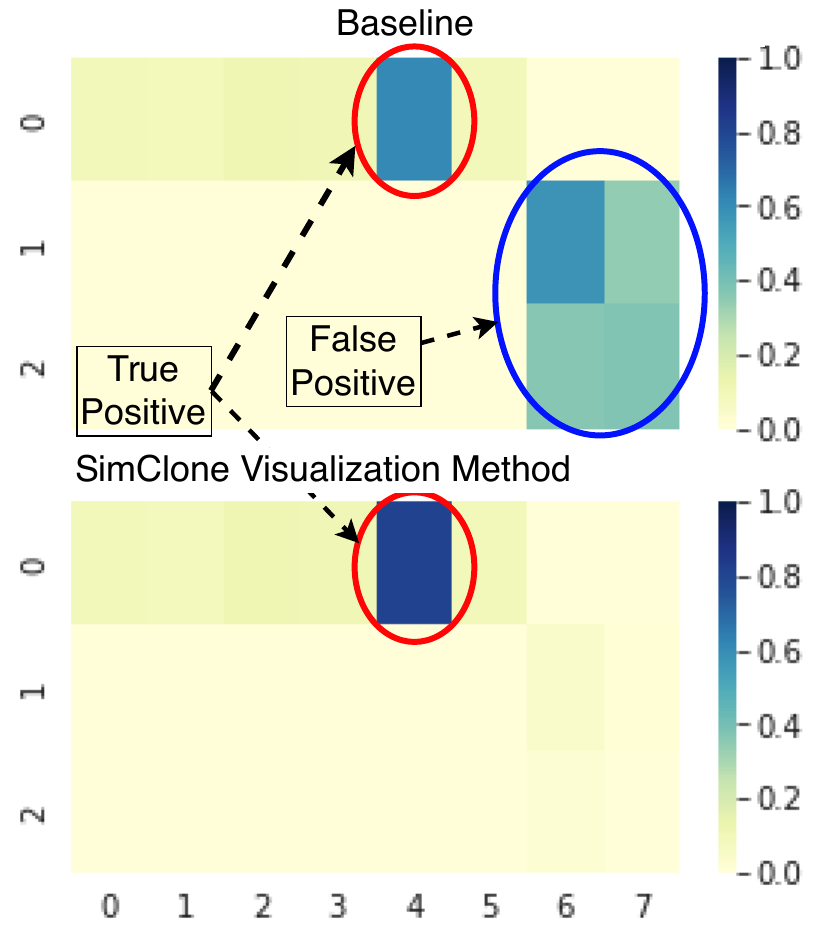}
    \caption{Example of SimClone Visualization Method helps with locating clones}
    \label{fig:shapviz_example}
\end{figure}

\begin{table}[H]
\caption{Comparison of our SimClone's visualization approach vs. baseline in terms of P@K (K = 1, 5, 10) when examining different numbers of top true \textit{ClonePairs} predicted by SimClone.}
%\resizebox{\columnwidth}{!}{%
\begin{tabular}{lllll}
\hline
\textbf{\#Prediction} & \textbf{Approach}      & \textbf{P@1}  & \textbf{P@5} & \textbf{P@10}  \\\hline
Top 20 & Our approach & \multicolumn{1}{r}{0.80} & \multicolumn{1}{r}{0.80} & \multicolumn{1}{r}{0.80}   \\
Top 20 & Baseline     & \multicolumn{1}{r}{0.40} & \multicolumn{1}{r}{0.40} & \multicolumn{1}{r}{0.40}  \\\hline
Top 50 & Our approach & \multicolumn{1}{r}{0.64} & \multicolumn{1}{r}{0.64} & \multicolumn{1}{r}{0.64}   \\
Top 50 & Baseline     & \multicolumn{1}{r}{0.54} & \multicolumn{1}{r}{0.53} & \multicolumn{1}{r}{0.53}  \\\hline
Top 100 & Our approach & \multicolumn{1}{r}{0.49} & \multicolumn{1}{r}{0.51} & \multicolumn{1}{r}{0.42}   \\
Top 100 & Baseline     & \multicolumn{1}{r}{0.42} & \multicolumn{1}{r}{0.42} & \multicolumn{1}{r}{0.42}  \\\hline
All & Our approach & \multicolumn{1}{r}{0.45} & \multicolumn{1}{r}{0.45} & \multicolumn{1}{r}{0.45}   \\
All & Baseline     & \multicolumn{1}{r}{0.31} & \multicolumn{1}{r}{0.31} & \multicolumn{1}{r}{0.30}  \\\hline
\end{tabular}
%}
\label{tab:RQ3}
\end{table}
\textbf{Our SimClone method's visualization outperforms the baseline method at least by 45\% on all true \textit{ClonePairs} in terms of Precision@K for all studied values of K.}
Table~\ref{tab:RQ3} compares our SimClone's visualization method's results with the baseline in terms P@K when examining different numbers of top predicted true \textit{ClonePairs}. In general, SimClone's visualization method outperforms baselines across all evaluation metrics. Figure\ref{fig:shapviz_example} shows an example where using SimClone method's visualization presents an advantage over using the baseline visualization method. We observe that due to the weighting of similarity values using the feature importance scores, we are able to avoid false positives whereas the baseline visualization method falls prey to them.

\textbf{Our visualization is more reliable for the predictions where the SimClone is confident.} If we compare the performance of our approach across the different number of top predictions (given by sorting the true positive \textit{ClonePairs} based on the probability given by SimClone), P@1 drops from 0.8 to 0.45 from the top 20 to all. The same trend can be observed for P@5 and P@10. This observation indicates that the visualization approach works more effectively when SimClone has higher confidence in its prediction. This observation is reasonable since our heatmap is based on the similarity matrices, which are in turn used as the feature after pooling. Better features result in a more discriminating classifier and hence predictions with more confidence.

\subsection{Results of RQ3: Effectiveness of Different Similarity Metrics}\label{sec:rq3}

\begin{table}[htbp]
  \centering
  \caption{The performance of approaches using different similarity matrices. SimClone Lite combines Jaccard, Textrank, Mean, and SD (ACC- Accuracy, P- Precision, R- Recall). For ease of comparison, we use the original SimClone as the baseline for time comparison and mark it as 1x in Table.}
    \begin{tabular}{m{6.32em}m{2.82em}lllll}\hline
    \textbf{Similarity Metric} & \textbf{Time}  & \textbf{ACC} & \textbf{AUC} & \textbf{F1} & \textbf{P} & \textbf{R} \\\hline
    all ($t$=10\%) & 1x    & \multicolumn{1}{r}{0.83}  & \multicolumn{1}{r}{0.91}  & \multicolumn{1}{r}{0.83}  & \multicolumn{1}{r}{0.85}  & \multicolumn{1}{r}{0.80} \\
    Jaccard & 0.183x & \multicolumn{1}{r}{0.75}  & \multicolumn{1}{r}{0.83}  & \multicolumn{1}{r}{0.77}  & \multicolumn{1}{r}{0.75}  & \multicolumn{1}{r}{0.78} \\
    Simhash & 0.128x & \multicolumn{1}{r}{0.59}  & \multicolumn{1}{r}{0.62}  & \multicolumn{1}{r}{0.47}  & \multicolumn{1}{r}{0.71}  & \multicolumn{1}{r}{0.35} \\
    Levenshtein & 0.392x & \multicolumn{1}{r}{0.61}  & \multicolumn{1}{r}{0.66}  & \multicolumn{1}{r}{0.49}  & \multicolumn{1}{r}{0.76}  & \multicolumn{1}{r}{0.36} \\
    Textrank & 0.158x & \multicolumn{1}{r}{0.63}  & \multicolumn{1}{r}{0.67}  & \multicolumn{1}{r}{0.53}  & \multicolumn{1}{r}{0.78}  & \multicolumn{1}{r}{0.40} \\
    Mean  & 0.099x & \multicolumn{1}{r}{0.64}  & \multicolumn{1}{r}{0.70}  & \multicolumn{1}{r}{0.59}  & \multicolumn{1}{r}{0.70}  & \multicolumn{1}{r}{0.51} \\
    SD    & 0.099x & \multicolumn{1}{r}{0.63}  & \multicolumn{1}{r}{0.69}  & \multicolumn{1}{r}{0.60}  & \multicolumn{1}{r}{0.67}  & \multicolumn{1}{r}{0.55} \\
    \rowcolor[rgb]{ .886,  .937,  .855}  SimClone Lite & 0.539x & \multicolumn{1}{r}{0.81}  & \multicolumn{1}{r}{0.90}  & \multicolumn{1}{r}{0.81}  & \multicolumn{1}{r}{0.82}  & \multicolumn{1}{r}{0.80} \\\hline
    \end{tabular}%
  \label{tab:time}%
\end{table}%

%SimClone uses six different similarity metrics. Even though we calculate similarities at row and column level instead of cell level to reduce SimClone's time complexity, such calculation can still be expensive. For $n$ datasets, similarity metrics of ${n \choose 2}$ pairs need to be calculated. In previous sections we examined the effectiveness of the metrics as a bundle. However, each metric may not contribute equally to the classifier. Therefore, identifying the most effective similarity metrics and reducing the number of similarity metrics can significantly improve the efficiency of SimClone. 

Table~\ref{tab:time} shows the result of the ablation analysis. We observed that when only Jaccard is used, the computation time could be reduced by 82\%, while F1-score suffers a slight drop from 0.83 to 0.77, compared to when all metrics are used. Among similarity metrics that only apply to strings, Textrank achieves better performance than Simhash and Levenshtein across all performance metrics, while having a relatively low computation time (0.158x). Both Mean and SD take very little time ($<$0.1x) to calculate. Therefore, we trained a lite version of SimClone (i.e., SimClone Lite) with a combination of Jaccard, Textrank, Mean, and SD. The result shows that we can shorten calculation time by 46\% and only lose 2\% of F1.

\subsection{Results of RQ4: Impact of Different Threshold $t$}\label{sec:rq4}

\textbf{In general, the effectiveness of both LTC and SimClone declines as the threshold $t$ increases, SimClone always outperforms LTC across different values of $t$.} Table~\ref{tab:RQ5} presents the performance of SimClone and LTC across different values of $t$. F1 of SimClone and LTC decreases from 0.863 to 0.745. F1 of LTC increases from 0.597 to 0.613 when $t$ increases from 0\% to 15\% and decreases after 15\%. This probably indicates that it is easier for both the methods to identify \textit{ClonePairs} when there is no or little noise (i.e., some amount of cloned data, which doesn't amount to qualifying as \textit{ClonePair}), even if the amount of noise is (e.g., 5\%) low. However, we note that our method SimClone is more robust to such noise compared to LTC as shown in Table~\ref{tab:RQ5}. Even when the $t$ is 90\% SimClone has an F1-score of 0745. while LTC only has an F1-score of 0.608.

\begin{table}

\caption{The performance of SimClone when using different thresholds $t$ for labeling. }\label{tab:RQ5}
% \resizebox{\columnwidth}{!}{%
\begin{tabular}{cllllll}
\hline
\textbf{Threshold $t$ }           & \textbf{Approach}  & \textbf{Accuracy} & \textbf{AUC} & \textbf{F1} & \textbf{Precision} & \textbf{Recall} \\ \hline

\multirow{2}{*}{0\%}          & SimClone & 0.869                        & 0.943                   & 0.863                  & 0.903                         & 0.826                      \\ 
                              & LTC      & 0.617                        & 0.673                   & 0.597                  & 0.631                         & 0.566                      \\\hline
\multirow{2}{*}{5\%}          & SimClone & 0.818                        & 0.901                   & 0.812                  & 0.808                         & 0.816                      \\
                              & LTC      & 0.620                        & 0.666                   & 0.612                  & 0.603                         & 0.622                      \\\hline
\multirow{2}{*}{10\%}         & SimClone & 0.853                        & 0.923                   & 0.851                  & 0.883                         & 0.820                      \\
                              & LTC      & 0.610                        & 0.655                   & 0.613                  & 0.620                         & 0.607                      \\\hline
\multirow{2}{*}{15\%}         & SimClone & 0.796                        & 0.885                   & 0.790                  & 0.794                         & 0.787                      \\
                              & LTC      & 0.586                        & 0.618                   & 0.574                  & 0.574                         & 0.575                      \\\hline
\multirow{2}{*}{20\%}         & SimClone & 0.817                        & 0.898                   & 0.814                  & 0.804                         & 0.825                      \\
                              & LTC      & 0.543                        & 0.576                   & 0.538                  & 0.553                         & 0.523                      \\\hline
\multirow{2}{*}{30\%}         & SimClone & 0.802                        & 0.900                   & 0.808                  & 0.797                         & 0.820                      \\
                              & LTC      & 0.543                        & 0.576                   & 0.538                  & 0.553                         & 0.523                      \\\hline
\multirow{2}{*}{50\%}         & SimClone & 0.779                        & 0.861                   & 0.788                  & 0.761                         & 0.816                      \\
                              & LTC      & 0.556                        & 0.575                   & 0.556                  & 0.559                         & 0.553                      \\\hline
\multirow{2}{*}{70\%}         & SimClone & 0.779                        & 0.861                   & 0.788                  & 0.761                         & 0.816                      \\
                              & LTC      & 0.740                        & 0.808                   & 0.745                  & 0.734                         & 0.756                      \\\hline
\multirow{2}{*}{90\%}         & SimClone & 0.740                        & 0.808                   & 0.745                  & 0.734                         & 0.756                      \\
                              & LTC      & 0.606                        & 0.624                   & 0.608                  & 0.607                         & 0.608              \\
                              \bottomrule
\end{tabular}
% }
\end{table}

\section{Discussion} \label{sec:discussion}

\subsection{Implications of our study and future research}\label{sec:implications}

\subsubsection{Data traceability.} Data clone detection is a crucial aspect of the Extract, Transform, Load (ETL) process as it aids in data traceability~\cite{lee2019big}. ETL refers to the process of integrating data from various sources into a single, consistent format before it is fed into a data warehouse or other destination system. However, during this process, the original format of the data may be altered, making it difficult to trace the data back to its source. Our proposed approach SimClone can be used to establish the connection between sourced data and destination data, ensuring the integrity and reliability of data. Future research is encouraged to leverage our data clone approach to facilitate data traceability in practice. 

\subsubsection{Data license copyright.} Many datasets have specific usage restrictions or requirements for attribution, and data clone detection enables organizations to follow these guidelines by detecting when a dataset has been used in the creation of another dataset~\cite{labastida2020licensing,khayyat2015open}. Data clone detection plays a critical role in ensuring compliance with data license copyright in tabular datasets. By identifying and flagging clone data, SimClone can assist organizations in ensuring that datasets are used in accordance with their associated licenses. This can aid organizations in avoiding legal liability for copyright infringement and help them ensure that they are using data in a legally compliant manner. Additionally, data clone detection can assist organizations in identifying and building the lineage relationship of the data, which is essential for proper attribution and adherence to usage restrictions. We encourage future research to investigate the effectiveness of SimClone in this field.

\subsubsection{Data Leakage in Machine Learning.} Data leakage refers to the phenomenon where information from the testing set leaks into the training set, leading to machine learning models that are not generalizable~\cite{hannun2021measuring}. One of the primary sources of target leakage is the presence of data clones, or multiple copies of the same data within a dataset~\cite{lopez2024inter}. These data clones can contain information about the testing set, while also present in the training set, leading to an overly optimistic model performance. Data clone detection plays a vital role in preventing data leakage in machine learning. By identifying and eliminating data clones from the training set using SimClone, future research can reduce the risk of target leakage and improve the generalizability and reliability of the trained model. %In addition to its role in preventing target leakage, data clone detection can also aid in the identification and correction of errors or inconsistencies in the data collection process. This can be accomplished by detecting duplicate or highly correlated data and removing it from the dataset, allowing organizations to ensure that they are using accurate and reliable data. 

\subsection{Limitations and potential solutions}
\subsubsection{Accelerating computation}
One limitation of SimClone is that it needs to calculate various similarity metrics, which is time-consuming. However, there are multiple ways to address the limitation. Firstly, as deliberated in RQ4, the impact of each similarity metric differs. For instance, only utilizing Jaccard could yield an F1 score of 0.77, compared to using all metrics, while consuming only 0.183 of the time required when using all metrics. Hence, practitioners could balance between efficiency and effectiveness according to their specific requirements. Second, If we look at the similarity computation in Section~\ref{sec:simmat}, we notice that the computation could be performed in parallel at different levels. The step of calculating different row-row/column-column similarities is parallelizable as well. In other words, the computation could be significantly accelerated by parallelization. One possible way to do this is to use a MapReduce framework which allows for the parallel processing of the data. Third, certain filters could be developed to reduce the number of dataset pairs. For instance, if a pair of datasets have different ranges of value, i.e., one ranging from 10 to 100, and another ranging from 0 to 1. Then there is no need to calculate similarity metrics, they are considered as non-clones given our definition of type 1 and type 3 clones. Our study is the first investigation in this direction and we encourage future research to improve our approach.

\section{Threats to Validity}\label{sec:threats}
In this section, we discuss the threats to the validity of our study.

\noindent\textbf{Internal validity}
One threat to internal validity exists when generating labels for the synthetic data. We labeled a data pair that has less than $t$ amount of injection as data clones. However, there may exist data clones in the dataset prior to clone injection, which may bias the evaluation results. To reduce the bias, we evaluated SimClone on a real-world dataset and confirmed that SimClone outperforms SOTA method. In a real-world scenario, practitioners probably consider the different amounts of duplicated data as clones. We keep it as a threshold that allows practitioners to tune and train the clone detection classifier with different degrees of sensitivity. In this study, we set $t$ to 10\% and use this setting across RQ1 to RQ4, which may pose a threat to our observations. To mitigate this threat, we investigated different values of $t$ in RQ5 and the results still hold.
Another threat exists in the manual label process of \textit{ClonePairs} in Section~\ref{sec:exp_data}. To mitigate the threat, the first two of the authors labeled the data independently and any disagreement was resolved, with a sufficient inter-rater agreement. Another threat is that in this study, we focus on using value-based similarities as features, and we did not include formatting attributes such as cell formula, font size, and font color as features in SimClone. We do so since large tabular datasets are typically stored in formats like CSV which do not contain such formatting attributes. However, our approach could be easily extended to include such formatting-related features. We encourage future research to combine both the value-based similarities and formatting attribution as features and examine their effectiveness
LTC paper does not provide implementation and labeled datasets. Therefore, in this study, we implemented LTC by ourselves, which is a potential validity. To mitigate the threat, we tried our best to carefully implement LTC to ensure its correctness and keep the same setting as they used.  

In RQ3, to evaluate the visualization component of SimClone, we use Precision@K to measure its accuracy, which could pose a threat to the validity of our study. Other evaluation methods could also be considered, such as a user study to evaluate the usefulness of our visualization component. We encourage future research in this direction.
%In addition, SimClone uses six different value-based similarities as features, there might be other more advanced similarity metrics to help improve our approach. However, this is the very first study investigating tabular clone detection, I encourage future research to investigate more features.

\noindent\textbf{External validity}
We evaluated SimClone on both a synthetic dataset and a real-world dataset. However, the results may not generalize to other datasets. We open-sourced all artifacts of this paper (i.e., code and datasets) to encourage future research to further verify SimClone, and construct new benchmarks for data clone detection.

%The external validity of our classifier may be threatened by the fact that it only learns and discriminates the existence of data clones between datasets by the value similarity between datasets with artificially injected data clones. In our experiments, we use a random clone injection ratio of 5-30\%, which may not exactly match the actual situation. 

%We performed sensitivity tests and tested the performance of our approach under different injection intervals. A smaller percentage of data clones degrades the classifier performance, as one would expect. However, we do not know the proportion of data clones that may exist in real-world datasets under different usage scenarios. Such different proportions of clones may result in pre-trained classifiers in some datasets that may not work in others.

\section{Conclusion}\label{sec:conclusion}
In this paper, we presented SimClone, the first data clone detection approach based on learning the value similarity between datasets. SimClone outperforms the existing SOTA method for detecting the data clone on a synthetic dataset that we create from UCI datasets repository and has high effectiveness on real-world datasets derived from the EUSES corpus. Furthermore, SimClone has a visualization method that helps the users of our method identify the exact location of data clones in a \textit{ClonePair}. In addition, we also provide a SimClone Lite, which executes in half the time as our original SimClone with only a slight drop in performance. 

Since our SimClone method does not require any formatting information and only relies on the similarity of the values in the dataset, SimClone can be used to detect data clones in multiple applications. For instance, one could use our SimClone method to identify if there are data clones between datasets with different licenses or copyrights and identify potential violations. Similarly, SimClone can be used to detect clones between test and train sets used to build AI software and identify if there are potential data leaks. Finally, SimClone can be very helpful in identifying the provenance and lineage information of different data points contained in a dataset (when compared against various sources) which is required for traceability purposes (e.g., to build an SBOM). 

\section*{Data Availability}
We make all the datasets, results and the code used in this study openly available in our replication package~\cite{simclonerepo}.

%%
%% The next two lines define the bibliography style to be used, and
%% the bibliography file.
\bibliographystyle{ACM-Reference-Format}
\bibliography{sample-base}

\end{document}